\documentclass[twocolumn, compsocconf]{IEEEtran}
\usepackage{cite}

\usepackage{mdwmath}
\usepackage{mdwtab}

\usepackage{url}
\usepackage[latin1]{inputenc}

\usepackage{multirow}

\usepackage{ifpdf}
\ifpdf\usepackage[raiselinks=false,colorlinks=true,citecolor=blue,urlcolor=blue,linkcolor=blue,bookmarksopen=true,hyperfootnotes=false,pdftex]{hyperref}\else
\usepackage[raiselinks=false,colorlinks=true,citecolor=blue,urlcolor=blue,linkcolor=blue,bookmarksopen=true,hyperfootnotes=false,dvips]{hyperref}\fi

\usepackage{amsthm}%Must be loaded before amsmath

\usepackage{setspace}
\usepackage{calc}
\usepackage{xspace}
\usepackage{verbatim}

\usepackage{rotating}
\usepackage{epsfig}

\usepackage{amssymb, bm}
\usepackage[cmex10]{amsmath}
\usepackage{mathabx}
\usepackage{array}
\usepackage{fixltx2e}

\usepackage{pstricks}
\usepackage{pst-plot}
\usepackage{pst-eucl}

\usepackage{mathrsfs}
\usepackage[only,shortrightarrow]{stmaryrd}

\usepackage[caption=false, font=footnotesize]{subfig}
\usepackage{url}

\newtheorem{theorem}{Theorem}[section]

\newtheorem{lemma}{Lemma}[section]

\newtheorem{definition}{Definition}[section]
\newcommand{\comp}{\mbox{\ensuremath{\mspace{2mu}\circ\mspace{2mu}}}}
\renewcommand{\leq}{\leqslant}
\renewcommand{\geq}{\geqslant}

\newcommand{\reduc}{\ensuremath{\psi}}%D\mspace{0mu}P}}}
\newcommand{\util}{\ensuremath{\mu}\xspace}

\newcommand{\dualMap}{\ensuremath{\varphi}}
\newcommand{\dual}{\ensuremath{\varphi}}

\newcommand{\pack}{\ensuremath{\pi}}
\newcommand{\vareps}{\ensuremath{\varepsilon}\xspace}
\newcommand{\Nat}{\ensuremath{\mathbb{N}}\xspace}
\newcommand{\Real}{\ensuremath{\mathbb{R}}\xspace}

\newcommand{\sched}{\ensuremath{\Sigma}\xspace}

\newcommand{\serv}{\ensuremath{\sigma}}
\newcommand{\clientOf}{\ensuremath{\Gamma}\xspace}
\newcommand{\futSet}{\ensuremath{\mathcal{T}}\xspace}
\newcommand{\servSet}{\ensuremath{\mathcal{T}}}
\newcommand{\jobSet}{\ensuremath{\mathcal{J}}}

\newcommand{\procSet}{\ensuremath{\Pi}\xspace}

\def\taskdef(#1,#2,#3){%
  \mbox{\ensuremath{\tau_{#1}\att(#2,#3)}}\xspace
}%
\newcommand{\att}{\ensuremath{\mspace{-5mu}:\mspace{-5mu}}}

\newlength{\ytrans}
\newlength{\ytmpc}

\newlength{\xtmpa}
\newlength{\ytmpa}
\newlength{\xtmpb}
\newlength{\ytmpb}
\newcommand{\subscr}{}

\newlength{\execWidth}

\newcounter{proc}
\newcounter{step}

\begin{document}

%% paper title
\title{
An Optimal Real-Time Scheduling Approach:\\From Multiprocessor to Uniprocessor
}

%Bridging the Gap between the Multiprocessors and Uniprocessor Real-Time Scheduling Problems

%Reduction of the Real-Time Tasks  Multiprocessor Scheduling Problem
%\\  to the Single Processor Scheduling Problem


\author{\IEEEauthorblockN{Paul Regnier, George Lima, Ernesto Massa\\}
  \IEEEauthorblockA{Computer Science Department --
    Distributed Systems Laboratory (LaSiD) --
    Federal University of Bahia, Brazil \\
    Email: \{pregnier, gmlima, ernestomassa\}@ufba.br}
% \and
% \IEEEauthorblockN{Authors Name/s per 2nd Affiliation (Author)}
% \IEEEauthorblockA{line 1 (of Affiliation): dept. name of organization\\
% line 2: name of organization, acronyms acceptable\\
% line 3: City, Country\\
% line 4: Email: name@xyz.com}
}


\maketitle

\begin{abstract}
  An optimal solution to the problem of scheduling real-time tasks on a set of
  identical processors is derived. The described approach is based on solving an
  equivalent uniprocessor real-time scheduling problem. Although there are other
  scheduling algorithms that achieve optimality, they usually impose prohibitive
  preemption costs. Unlike these algorithms, it is observed through simulation
  that the proposed approach produces no more than three preemptions points per
  job.
  % Unlike these algorithms, it is shown by simulation that the proposed
  % approach produces a few preemption points per job on average.
\end{abstract}


% \fi
\IEEEpeerreviewmaketitle

\section{Introduction}\label{sec:introduction}

\subsection{Motivation}\label{sec:motiv}

Scheduling $n$ real-time tasks on $m$ processors is a problem that has taken
considerable attention in the last decade. The goal is to find a feasible
schedule for these tasks, that is a schedule according to which no task misses
its deadlines. Several versions of this problem have been addressed and a number
of different solutions have been given. One of the simplest versions assumes a
periodic-preemptive-independent task model with implicit deadlines, PPID for
short. According to the PPID model each task is independent of the others, jobs
of the same task are released periodically, each job of a task must finish
before the release time of its successor job, and the system is fully
preemptive.
 
A scheduling algorithm is considered optimal if it is able to find a feasible
schedule whenever one exists. Some optimal scheduling algorithms for the PPID
model have been found. For example, it has been shown that if all tasks share
the same deadline \cite{McNaughton59}, the system can be optimally scheduled
with a very low implementation cost. The assumed restriction on task deadlines,
however, prevents the applicability of this approach. Other optimal algorithms
remove this restriction but impose a high implementation cost due to the
required number of task preemptions \cite{Baruah96, Cho06, Levin2010}. It is
also possible to find trade-offs between optimality and preemption cost
\cite{Andersson08, Massa10, Andersson08b, Bletsas09}.

Optimal solutions for the scheduling problem in the PPID model are able to
create preemption points that make it possible task migrations between
processors allowing for the full utilization of the system. As illustration
consider that there are three tasks, $\tau_1$, $\tau_2$ and $\tau_3$, to be
scheduled on two processors. Suppose that each of these tasks requires $2$ time
units of processor and must finish $3$ time units after they are released.
Also, assume that all three tasks have the same release time. As can be seen in
Figure \ref{fig:probSched}, if two of these tasks are chosen to execute at their
release time and they are not preempted, the pending task will miss its
deadline. As all tasks share the same deadline in this example, the approach by
McNaughton \cite{McNaughton59} can be applied, as illustrated in the figure. If
this was not the case, generating possibly infinitely many preemption points
could be a solution as it is shown by other approaches \cite{Baruah96, Cho06,
  Levin2010}. In this work we are interested in a more flexible solution.

% mudar A,B,C para \tau
\setlength{\execWidth}{0.37\psyunit}

\def\schedAxes(#1,#2){%
  \setcounter{step}{#2}%
  \multido{\nl=0+1}{\thestep}{%increments \na in step of 2 5 times
    \rput(\nl,-.25){\footnotesize \nl}%
  }%
  \multido{\i=0+1}{#1}{%
    \setcounter{proc}{\i+1}%
    \setlength{\ytmpa}{\i\psyunit+.14\psyunit}%
%    \rput(-0.5,\ytmpa){\small $\pi_{\theproc}$}%
    \setlength{\ytmpa}{\i\psyunit}%
    \setlength{\ytmpb}{\i\psyunit-.07\psyunit}%
    \setlength{\xtmpa}{#2\psxunit-0.5\psxunit}% 
    \psline[linewidth=0.5pt]{->}(0,\ytmpa)(\xtmpa,\ytmpa)%
    \multido{\nt=0+1}{#2}{
      \psline[linestyle=solid,linewidth=0.5pt](\nt,\ytmpa)(\nt,\ytmpb)%
    }%
  }%
}%
\def\jobLocExec#1#2(#3,#4,#5){%
  \setcounter{proc}{#5-1}%
  \setlength{\xtmpa}{#3\psxunit}%
  \setlength{\ytmpa}{\theproc\psyunit}%
  \setlength{\xtmpb}{#3\psxunit+#4\psxunit}%
  \setlength{\ytmpb}{\theproc\psyunit+\execWidth}%
  \psset{linecolor=black,linestyle=solid}%
  \psframe[linecolor=black,linestyle=solid,fillstyle=none]
  (\xtmpa,\ytmpa)(\xtmpb,\ytmpb)%
  \ifthenelse{\equal{#1}{}}{}{%
  }{%
    \setlength{\xtmpa}{#3\psxunit+{#4\psxunit*\real{0.5}}}%
    \setlength{\ytmpb}{\theproc\psyunit+0.5\execWidth}%
    \ifthenelse{\equal{#2}{}}{%
      \renewcommand{\subscr}{#1}%
    }{%
      \renewcommand{\subscr}{#1$^{#2}$}%
    }%
    \rput(\xtmpa,\ytmpb){\footnotesize \subscr}%
  }%
}%
%#1: #task, #2: abs deadline, #3: #processor, #4: color
\def\jobdeadline#1(#2,#3){%
  \setcounter{proc}{#3-1}%
  \setlength{\xtmpa}{#2\psxunit}%
  \setlength{\ytmpa}{\theproc\psyunit}%
  \setlength{\ytmpb}{\theproc\psyunit+\execWidth+0.35\psyunit}%
  \psline[linestyle=solid,linewidth=0.5pt,arrows=-*,arrowsize=4pt]
  (\xtmpa,\ytmpa)(\xtmpa,\ytmpb)%
  \ifthenelse{\equal{#1}{}}{}{%
    \setlength{\ytmpa}{\theproc\psyunit}%
    \uput{.3em}[270](\xtmpa,\ytmpa){\vphantom{$d_j$}\footnotesize #1}}%
}%
\begin{figure}[tbh]
  \centering%
  \subfloat[C misses its deadline]{%
    \psset{xunit=0.8cm, yunit=1cm}%
    \begin{pspicture*}(-.2,-0.4)(4.,1.8)%
      \schedAxes(2,4)%
      % \rput(-0.74,.14\psyunit){\small $\Sigma_1$}%
      % \rput(-0.74,1.14\psyunit){\small $\Sigma_2$}%
      % Task 1
      \jobLocExec{$\tau_1$}(0, 2, 1)%
      % \jobexec{1}(6, 2, 1)%
      % Task 2
      \jobLocExec{$\tau_2$}(0, 2, 2)%
      % Task 3
      \jobLocExec{$\tau_3$}(2, 1, 1)%
      \jobdeadline{}(3,2)%
      \jobdeadline{}(3,1)%
    \end{pspicture*}
  }%
  \hspace{.1cm}% 
   \subfloat[Correct schedule]{%
    \psset{xunit=0.8cm, yunit=1cm}%
    \begin{pspicture*}(-.2,-0.4)(4.,1.8)%
      \schedAxes(2,4)%
      % Task 1
      \jobLocExec{$\tau_3$}(0, 2, 1)%
      % Task 2
      \jobLocExec{$\tau_1$}(1, 2, 2)%
      \jobLocExec{$\tau_2$}(0, 1, 2)%
      \jobLocExec{$\tau_2$}(2, 1, 1)%
      \jobdeadline{}(3,2)%
      \jobdeadline{}(3,1)%
      % Task 3
     \end{pspicture*}
   }%
   \caption{ A deadline miss occurs in case (a), but not in case (b).}
   \label{fig:probSched}
\end{figure}
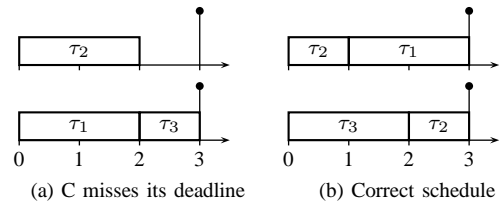

\subsection{Contribution}

% The proposed approach is able to optimally schedule a set of tasks on a
% multiprocessor system. Indeed, tasks are not required to be periodic. We assume
% that tasks utilize a fixed processor bandwidth within the interval between their
% respective release times and deadlines. For example, a job of a task using $u\%$
% of processor requires $u(d-r)$ execution time, where $r$ and $d$ represent its
% release time and deadline. We call this kind of task fixed-utilization
% task. Note that according to the PPID model, the value $d-r$ is the same for all
% jobs of the same task, which makes the model assumed in this paper slightly more
% general than the PPID model.

In the present work, we define a real-time task as an infinite sequence of
jobs. Each job represents a piece of work to be executed on one or more
processors. A job is characterized by its release time $r$, time after which it
can be executed, and its deadline $d$, time by which it must be completed in
order for the system to be correct. Also, we assume that the deadline of a job
is equal to the release time of the next job of the same task. However,
differently from the PPID model, we do not assume that tasks are necessarily
periodic. Instead, we assume that tasks have a fixed-utilization, i. e. each job
of a task utilizes a fixed processor bandwidth within the interval between its
release time and deadline. For example, a job of a task with utilization $u \leq
1$ of processor requires $u (d-r)$ execution time. Note that according to the
PPID model, the value $d-r$ is equal to the period of the periodic task, which
makes the model assumed in this paper slightly more general than the PPID model.

The proposed approach is able to optimally schedule a set of fixed-utilization
tasks on a multiprocessor system. The solution we describe does
not impose further restrictions on the task model and only a few preemption
points per job are generated. The idea is to reduce the real-time multiprocessor
scheduling problem into an equivalent real-time uniprocessor scheduling
problem. After solving the latter, the found solution is transformed back to a
solution to the original problem. This approach seems very attractive since it
makes use of well known results for scheduling uniprocessor systems.

\begin{figure}[tbh]
  \centering%
  \label{fig:validSched}
  \centering%
  \setlength{\execWidth}{0.42\psyunit}

  \def\schedAxes(#1,#2){%
    \setcounter{step}{#2}%
    \multido{\nl=0+1}{\thestep}{%increments \na in step of 2 5 times
      \rput(\nl,-.25){\footnotesize \nl}%
    }%
    \multido{\i=0+1}{#1}{%
      \setcounter{proc}{\i+1}%
      \setlength{\ytmpa}{\i\psyunit+.14\psyunit}%
      % \rput(-0.5,\ytmpa){\small $\pi_{\theproc}$}%
      \setlength{\ytmpa}{\i\psyunit}%
      \setlength{\ytmpb}{\i\psyunit-.07\psyunit}%
      \setlength{\xtmpa}{#2\psxunit-0.5\psxunit}%
      \psline[linewidth=0.5pt]{->}(0,\ytmpa)(\xtmpa,\ytmpa)%
      \multido{\nt=0+1}{#2}{
        \psline[linestyle=solid,linewidth=0.5pt](\nt,\ytmpa)(\nt,\ytmpb)%
      }%
    }%
  }%
  \def\jobLocExec#1#2(#3,#4,#5){%
    \setcounter{proc}{#5-1}%
    \setlength{\xtmpa}{#3\psxunit}%
    \setlength{\ytmpa}{\theproc\psyunit}%
    \setlength{\xtmpb}{#3\psxunit+#4\psxunit}%
    \setlength{\ytmpb}{\theproc\psyunit+\execWidth}%
    \psset{linecolor=black,linestyle=solid}%
    \psframe[linecolor=black,linestyle=solid,fillstyle=none]
    (\xtmpa,\ytmpa)(\xtmpb,\ytmpb)%
    \ifthenelse{\equal{#1}{}}{}{%
    }{%
      \setlength{\xtmpa}{#3\psxunit+{#4\psxunit*\real{0.5}}}%
      \setlength{\ytmpb}{\theproc\psyunit+0.5\execWidth}%
      \ifthenelse{\equal{#2}{}}{%
        \renewcommand{\subscr}{#1}%
      }{%
        \renewcommand{\subscr}{#1$^{#2}$}%
      }%
      \rput(\xtmpa,\ytmpb){\footnotesize \subscr}%
    }%
  }%
  % #1: #task, #2: abs deadline, #3: #processor, #4: color
  \def\jobdeadline#1(#2,#3){%
    \setcounter{proc}{#3-1}%
    \setlength{\xtmpa}{#2\psxunit}%
    \setlength{\ytmpa}{\theproc\psyunit}%
    \setlength{\ytmpb}{\theproc\psyunit+\execWidth+0.35\psyunit}%
    \psline[linestyle=solid,linewidth=0.5pt,arrows=-*,arrowsize=4pt]
    (\xtmpa,\ytmpa)(\xtmpa,\ytmpb)%
    \ifthenelse{\equal{#1}{}}{}{%
      \setlength{\ytmpa}{\theproc\psyunit}%
      \uput{.3em}[270](\xtmpa,\ytmpa){\vphantom{$d_j$}\footnotesize #1}}%
  }%
  \psset{xunit=1.cm, yunit=1.cm}%
  \begin{pspicture*}(-1.2,-0.4)(4.,3)%
    \schedAxes(3,4)%
    % dual
    % \uput{.3em}[270](-0.8,1.36){\vphantom{$d_j$}Schedule of $A^*$, $B^*$ and
    % $C^*$}%
    % \uput{.3em}[270](-0.8,0.36){\vphantom{$d_j$}$\Sigma_2$}%
    % \uput{.3em}[270](-0.8,2.36){\vphantom{$d_j$}$\Sigma^*$}%
     % 
    \jobLocExec{$\tau_1^*$}(0, 1, 3)%
    \jobLocExec{$\tau_2^*$}(1, 1, 3)%
    \jobLocExec{$\tau_3^*$}(2, 1, 3)%

    \jobLocExec{$\tau_3$}(0, 2, 1)%
    \jobLocExec{$\tau_2$}(0, 1, 2)%
    % Task 2
    \jobLocExec{$\tau_2$}(2, 1, 1)%
    \jobLocExec{$\tau_1$}(1, 2, 2)%
    % Task 3
    \jobdeadline{}(3,3)%
    \jobdeadline{}(3,2)%
    \jobdeadline{}(3,1)%
  \end{pspicture*}
  \caption{Scheduling equivalence of $\tau_1^*$, $\tau_2^*$ $\tau_3^*$ on one
    processor and $\tau_1$, $\tau_2$, $\tau_3$ on two processors.}
  \label{fig:dualSchedEx}
\end{figure}
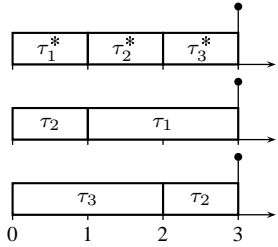

Consider the illustrative system with 3-tasks previously given. We show that
scheduling this system on two processors is equivalent to scheduling another
3-task system with tasks $\tau_1^*$, $\tau_2^*$ and $\tau_3^*$ on one
processor. Each star task requires one unit of time and has the same deadline as
the original task, that is the star tasks represent the slack of the original
ones. As can be seen in Figure \ref{fig:dualSchedEx}, the basic scheduling rule
is the following. Whenever the star task executes on the transformed system, its
associated original task does not execute on the original system. For example,
when $\tau_1^*$ is executing on the transformed system, task $\tau_1$ is not
executing on the original system.

The illustrative example gives only a glimpse of the proposed approach and does
not capture the powerfulness of the solution described in this document. For
example, if the illustrative example had four tasks instead of three, the
scheduling rule could not be applied straightforwardly. For such cases, we show
how to aggregate tasks so that the reduction to the uniprocessor scheduling
problem is still possible. For more general cases, a series of system
transformation, each one generating a system with fewer processors, may be
applied. Once a system with only one processor is obtained, the well known EDF
algorithm is used to generate the correct schedule. Then, it is shown that this
schedule can be used to correctly generate the schedule for the original
multiprocessor system.

\subsection{Structure}

In the remainder of this paper we detail the proposed approach. The notation and
the assumed model of computation are described in Section
\ref{sec:model}. Section \ref{sec:server} presents the concept of servers, which
are a means to aggregate tasks (or servers) into a single entity to be
scheduled. In Section \ref{sec:virtualSched} it is shown the rules to transform
a multiprocessor system into an equivalent one with fewer processors and the
scheduling rules used. The correctness of the approach is also shown in this
section. Then, experimental results collected by simulations are presented in
Section \ref{sec:evaluation}. Finally, Section \ref{sec:relatedWork} gives a
brief summary on related work and conclusions are drawn in Section
\ref{sec:conclusion}.

\section{System Model and Notation}\label{sec:model}

\subsection{Fixed-Utilization Tasks}\label{sec:fixedUtilTask}

As mentioned earlier, we consider a system comprised of $n$ real-time and
independent tasks, each of which defines an infinite sequence of released
jobs. More generally, a job can be defined as follows.

\begin{definition}[Job]
  A real-time job, or simply, job, is a finite sequence of instructions to be
  executed. If $J$ is a job, it admits a release time, denoted $J.r$, an
  execution requirement, denoted $J.c$, and a deadline, denoted $J.d$.
\end{definition}

In order to represent possibly non-periodic execution requirements, we introduce
a general real-time object, called fixed-utilization task, or task for short,
whose execution requirement is specified in terms of processor utilization
within a given interval. Since a task shall be able to execute on a single
processor, its utilization cannot be greater than one.

\begin{definition}[Fixed-Utilization Task]\label{dfn:fixedUtilTask}
  Let $u$ be a positive real not greater than one and let $D$ be a countable and
  unbounded set of non-negative reals. The fixed-utilization task $\tau$ with
  utilization $u$ and deadline set $D$, denoted $\taskdef(,u,D)$, satisfies the
  following properties: (i) a job of $\tau$ is released at time $t$ if and only
  if $t \in D$; (ii) if $J$ is released at time $r$, then $J.d = \min_{t} \{ t
  \in D, t > J.r \}$; and (iii) $J.c = u (J.d - J.r)$.
\end{definition}

Given a fixed-utilization task $\tau$, we denote $\util(\tau)$ and
$\Lambda(\tau)$ its utilization and its deadline set, respectively.
% \begin{definition}[Fixed-Utilization Task]\label{dfn:fixedUtilTask}
%   A fixed-utilization task $\,\tau$ with start time $s$ and utilization
%   $\util(\tau) \leqslant 1\,$ is an infinite sequence of jobs such that: (i) the
%   first job of $\tau$ is released at $s$; (ii) if $J$ and $J'$ are
%   two consecutive jobs of $\tau$, then $J.d = J'.r$; and (iii) for any job
%   $J$ of $\tau$, $J.c = \util(\tau) (J.d - J.r)$. The set of all deadlines of a
%   fixed-utilization task $\tau$ is denoted as $\Lambda(\tau)$.
% \end{definition}

% We denote $\taskdef(,u, \Lambda)$ a fixed utilization task $\tau$ of
%   utilization $u$ and set of deadlines $\Lambda$.

As a simple example of fixed-utilization task, consider a periodic task $\tau$
characterized by three attributes: (i) its start time $\,s$; (ii) its period
$T$; and (iii) its execution requirement $C$. Task $\tau$ generates an infinite
collection of jobs each of which released at $s + (j-1) T$ and with deadline at
$s + j T$, $j \in \Nat^*$. Hence, $\tau$ can be seen as a fixed-utilization task
with start time at $s$, utilization $\util(\tau) = C/T$ and set of deadlines
$\Lambda(\tau) = \{ (s + j T), j \in \Nat^* \}$, which requires exactly
$\util(\tau) T$ of processor during periodic time intervals $[s + (j-1) T, s + j
T)$, for $j$ in $\in \Nat^*$.  As will be clearer later on, the concept of
fixed-utilization task will be useful to represent non-periodic processing
requirements, such as those required by groups of real-time periodic tasks.

% For the sake of conciseness, we simply call task a fixed-utilization task
%   when no confusion is introduced doing so.

\subsection{Fully Utilized System}
\label{sec:fullyUtilSyst}

We say that a set of $n$ fixed-utilization tasks fully utilizes a system
comprised of $m$ identical processors if the sum of the utilizations of the $n$
tasks exactly equals $m$. Hereafter, we assume that the set of $n$
fixed-utilization tasks fully utilizes the system.

It is important to mention that this assumption does not restrict the
applicability of the proposed approach.  For example, if a job $J$ of a task is
supposed to require $J.c$ time units of processor but it completes consuming
only $c'< J.c$ processor units, then the system can easily simulate $J.c - c'$
of its execution by blocking a processor accordingly. Also, if the maximum
processor utilization required by the task set is less than $m$, dummy tasks can
be created to comply with the full utilization assumption. Therefore, we
consider hereafter that the full utilization assumption holds and so each job
$J$ executes exactly for $u(J.d-J.r)$ time units during $[r,d)$.

% Although the synchronous and the periodic assumption can be relaxed, allowing
% to apply our approach to asynchronous sporadic job sets, such results are out
% of the scope of this seminal work.

\subsection{Global Scheduling} 

Jobs are assumed to be enqueued in a global queue and are scheduled to execute
on a multiprocessor platform $\procSet$, comprised of $m > 1$ identical
processors. We consider a global scheduling policy according to which tasks are
independent, preemptive and can migrate from a processor to another during their
executions. There is no penalty associated with preemptions or migrations.

\begin{definition}[Schedule]
  For any collection of jobs, denoted $\jobSet$, and multiprocessor platform
  $\procSet$, the multiprocessor schedule $\sched$ is a mapping from $\,\Real^+
  \times \jobSet \times \procSet$ to $\{0, 1\}$ with $\sched(t, J, \pi)$ equal
  to one if schedule $\sched$ assigns job $J$ to execute on processor $\pi$ at
  time $t$, and zero otherwise.
\end{definition}

Note that by the above definition, the execution requirement of a job $J$ at
time $t$ can be expressed as
\[
e(J,t) = J.c - \sum_{\pi \in \procSet} \int_{J.r}^t \sched(t, J, \pi) dt,
\]

\begin{definition}[Valid Schedule]
  A schedule $\,\sched$ of a job set $\jobSet$ is valid if (i) at any time, a
  single processor executes at most one job in $\jobSet$; (ii) any job in
  $\jobSet$ does not execute on more than one processor at any time; (iii) any
  job $J \in \jobSet$ can only execute at time $t$ if $J.r \leq t$ and $e(J,t) >
  0$.
\end{definition}

\begin{definition}[Feasible Schedule]
  Let $\;\sched$ be a schedule of a set of jobs $\;\jobSet$. The schedule
  $\sched$ is feasible if it is a valid schedule and if all the jobs in
  $\jobSet$ finish executing by their deadlines.
\end{definition}

We say that a job is feasible in a schedule $\sched$ if it finishes executing by
its deadline, independently of the feasibility of $\sched$. That is, a job can
be feasible in a non-feasible schedule. However, if $\sched$ is feasible, then
all jobs scheduled in $\sched$ are necessarily feasible. Also, we say that a job
$J$ is active at time $t$ if $J.r \leq t$ and $e(J,t) > 0$.  As a
consequence, a fixed-utilization task admits a unique feasible and active job at
any time.

% Note that a job of a task can execute less than its worst-case execution time
% and still be feasible. However, this is not the case for a job of fixed
% utilization must always take its execution time requirement by its deadline
% according to Definition \ref{dfn:fixedUtilTask}.

% In this work we are interested in work-conserving scheduling algorithms. 
% \begin{definition}[Work-Conserving]
%   A scheduling algorithm is work-conserving if it never idles a processor
%   whenever some job is ready to execute.
% \end{definition}

\section{Servers}\label{sec:server}

As mentioned before, the derivation of a schedule for a multiprocessor system will
be done via generating a schedule for an equivalent uniprocessor system. One of
the tools for accomplishing this goal is to aggregate tasks into servers, which
can be seen as fixed-utilization tasks equipped with a scheduling mechanism.

As will be seen, the utilization of a server is not greater than one. Hence, in
this section we will not deal with the multiprocessor scheduling problem. The
focus here is on precisely defining the concept of servers (Section
\ref{sec:serverModel}) and showing how they correctly schedule the
fixed-utilization tasks associated to them (Section
\ref{sec:predictability}). In other words, the reader can assume in this section
that there is a single processor in the system. Later on we will show how
multiple servers are scheduled on a multiprocessor system.

\subsection{Server model and notations}
\label{sec:serverModel}

A fixed-utilization server associated to a set of fixed-utilization tasks is
defined as follows:

% Suppose $0 < \util(\futSet) \leq 1$. The aggregation mechanisms consists in
% entrusting a fixed utilization task of utilization $\util(\futSet)$ to
% schedule the elements of $\futSet$. For this purpose, such a fixed utilization 
% task, called server, is equipped with a work-conserving scheduling policy.

\begin{definition}[Fixed-Utilization Server]\label{dfn:server}
  Let $\futSet$ be a set of fixed-utilization tasks with total utilization given
  by
  \[
  \util(\futSet) = \sum_{\tau \in \futSet} \util(\tau) \leqslant 1
  \]
  A fixed-utilization server $S$ associated to $\futSet$, denoted
  $\serv(\servSet)$, is a fixed-utilization task with utilization
  $\util(\futSet)$, set of deadlines $\Lambda(S) \subseteq \bigcup_{\tau \in
    \futSet} \Lambda(\tau)$, equipped with a scheduling policy used to schedule
  the jobs of the elements in $\futSet$. For any time interval $[d,d')$, where
  $d,d' \in \Lambda(S)$, $S$ is allowed to execute exactly for
  $\util(\futSet)(d'-d)$ time units.
\end{definition}

% obs-paul: melhorar o operador
Given a fixed-utilization server $S$, we denote $\clientOf(S)$ the set of
fixed-utilization tasks scheduled by $S$ and we assume that this set is
statically defined before the system execution. Hence, the utilization of a
server, simply denoted $\util(S)$, can be consistently defined as equal to
$\util(\clientOf(S))$. Note that, since servers are fixed-utilization tasks, we
are in condition to define the server of a set of servers. For the of sake of
conciseness, we call an element of $\clientOf(S)$ a client task of $S$ and we
call a job of a client task of $S$ a client job of $S$.  If $S$ is a server and
$\servSet$ a set of servers, then $\serv(\clientOf(S)) = S$ and
$\clientOf(\serv(\servSet)) = \servSet$.

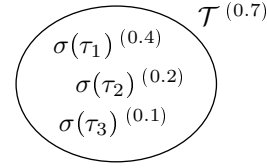
\begin{figure}[t]
  \centering%
  \psset{xunit=0.74cm, yunit=0.74cm}%
  \begin{pspicture*}(0,-0.4)(7,3)%
    \psset{fillstyle=none,linewidth=.5pt}%
    % \psgrid[subgriddiv=1,gridwidth=.5pt,griddots=4,subgriddots=0,gridlabels=.3]
    % (0,0)(10,10)%
    \psellipse[fillcolor=lightgray](2.5,1.2)(1.8,1.4)%
    \uput{0}[0](3.8,2.4){\vphantom{$d_j$} $\servSet^{\;(0.7)}$}%
    \uput{0}[0](1.2,1.9){\vphantom{$d_j$} $\serv(\tau_1)^{\;(0.4)}$}%
    \uput{0}[0](1.6,1.2){\vphantom{$d_j$} $\serv(\tau_2)^{\;(0.2)}$}%
    \uput{0}[0](1.3,0.5){\vphantom{$d_j$} $\serv(\tau_3)^{\;(0.1)}$}%
  \end{pspicture*}%

  \caption{A three-server set. The utilization $u$ of a server $S$ or a set of
    server $\servSet$ is indicated by the notation $S^{(u)}$ and
    $\servSet^{(u)}$, respectively. \label{fig:servSet}}
\end{figure}

For illustration consider Figure \ref{fig:servSet}, where $\servSet$ is a set
comprised of the three servers $\serv(\tau_1)$, $\serv(\tau_2)$ and
$\serv(\tau_3)$ associated to the fixed-utilization tasks $\tau_1$, $\tau_2$ and
$\tau_3$, respectively. The numbers between brackets represent processor
utilizations. If $S = \serv(\servSet)$ is the server in charge of scheduling
$\serv(\tau_1)$, $\serv(\tau_2)$ and $\serv(\tau_3)$, then we have $\servSet =
\clientOf(S) = \{\serv(\tau_1), \serv(\tau_2), \serv(\tau_3) \}$ and $\util(S) =
0.7$.

% Let $\Omega = \bigcup_{\tau \in \futSet} \Lambda(\tau)$. As can be seen by
% Definition \ref{dfn:server}, the server $S$ associated to $\futSet$ may not have
% all the elements of $\Omega$ since $\Lambda(S) \subseteq \Omega$.

As can be seen by Definition \ref{dfn:server}, the server $S$ associated to
$\futSet$ may not have all the elements of $\bigcup_{\tau \in \futSet}
\Lambda(\tau)$. Indeed, the number of elements in $\Lambda(S)$ depends on a
server deadline assignment policy:

\begin{definition}[Server Deadline Assignment\label{dfn:deadlineAssignt}]
  A deadline of a server $S$ at time $t$, denoted $\lambda_S(t)$, is given by
  the earliest deadline greater than $t$ among all client jobs of $S$ not yet
  completed at time $t$. This includes those jobs active at $t$ or the not yet
  released jobs at $t$. More formally,
  \[
  \lambda_S(t) = \min_{J \in \jobSet} \{ J.d, ( J.r < t \land e(J,t) > 0 ) \;
  \lor \; J.r \geqslant t\}
  \]
  where $\jobSet$ is the set of all jobs of servers in $\clientOf(S)$.
\end{definition}

Note that by Definitions \ref{dfn:fixedUtilTask} and \ref{dfn:deadlineAssignt},
the execution requirement of a server $S$ in any interval $(d,d')$ equals
$\util(S)(d'-d)$, where $d$ and $d'$ are two consecutive deadlines in
$\Lambda(S)$. As a consequence, the execution requirement of a job $J$ of a
server $S$, released at time $d \in \Lambda(S)$, equals $J.c = e(J,d) =
\util(S)(\lambda_S(d) - d)$ for all $d \in \Lambda(S)$. The budget of $S$ at any
time $t$, denoted as $C_S(t)$, is replenished to $e(J,t)$ at all $t \in
\Lambda(S)$. The budget of a server represents the processing time available for
its clients. Although a server never executes itself, we say that a server $S$
is executing at time $t$ in the sense that one of its client tasks consumes
its budget $C_S(t)$ at the same rate of its execution.

Recall from Section \ref{sec:fixedUtilTask} that a job of an fixed-utilization
task is feasible in a schedule $\sched$ if it meets its deadline. However, the
feasibility of a server does not imply the feasibility of its client tasks. For
example, consider two periodic tasks $\taskdef(1,1/2,2\Nat^*)$ and
$\taskdef(2,1/3, 3\Nat^* )$, with periods equal to 2 and 3 and
utilizations $\mu(\tau_1) = 1/2$ and $\mu(\tau_2) = 1/3$, respectively. Assume
that their start times are equal to zero. Consider a server $S$ scheduling these
two tasks on a dedicated processor and let $\Lambda(S) = \{2, 3, 4, 6, \ldots
\}$. Thus, the budget of $S$ during $[0,2)$ equals $C_S(0) = 2 \util(S) =
5/3$. Let $\sched$ be a schedule of $\tau_1$ and $\tau_2$ in which $S$ is
feasible.  The feasibility of server $S$ implies that $S$ acquires the processor
for at least $5/3$ units of time during $[0,2)$, since $2$ is a deadline of
$S$. Now, suppose that the scheduling policy used by $S$ to schedule its client
tasks gives higher priority to $\tau_2$ at time $0$.  Then, $\tau_2$ will
consume one unit of time before $\tau_1$ begins its execution. Therefore, the
remaining budget $C_S(1) = 2/3$ will be insufficient to complete $\tau_1$ by
$2$, its deadline. This illustrates that a server can be feasible while the
generated schedule of its clients is not feasible.

\subsection{EDF Server}
\label{sec:predictability}

In this section, we define an EDF server and shows that EDF servers are
predictable in the following sense.

\begin{definition}[Predictable Server]\label{dfn:predictableServer}
  A fixed-utilization server $S$ is predictable in a schedule $\sched$ if its
  feasibility in $\sched$ implies the feasibility of all its client jobs.
\end{definition}

\begin{definition}[EDF Server]\label{dfn:edfServer}
  An EDF server is a fixed-utilization server $S$, defined according to
  Definitions \ref{dfn:server} and \ref{dfn:deadlineAssignt}, which schedules
  its client tasks by EDF.
\end{definition}

For illustration, consider a set of three periodic tasks $\futSet =
\{\taskdef(1,1/3, 3\Nat^* ), \taskdef(2,1/4, 4\Nat^* ), \taskdef(3,1/6, 6\Nat^*
)\}$.  Since $\util(\futSet) = 3/4 \leqslant 1$, we can define an EDF server $S$
to schedule $\futSet$ such that $\clientOf(S) = \futSet$ and $\util(S) =
3/4$. Figure \ref{fig:budget} shows both the evolution of $C_S(t)$ during
interval $[0,12)$ and the schedule $\sched$ of $\futSet$ by $S$ on a single
processor. In this figure, $i_j$ represents the $j$-th job of $\tau_i$. Observe
here that $\Lambda(S) \neq \{ 3k, 4k, 6k | k \in \Nat^*\}$. Indeed, deadlines
$4$ of $2_1$ and $9$ of $1_3$ are not in $\Lambda(S)$, since $2_1$ and $1_3$ are
completed at time $3$ and $8$, respectively.

It is worth noticing that the deadline set of a server could be defined to
include all deadlines of its clients. However, this would generate unnecessary
preemption points.

\begin{figure}[t]
  \centering%
\setlength{\execWidth}{0.37\psyunit}

\def\schedAxes(#1,#2){%
  \setcounter{step}{((#2+1)/2)}%
  \multido{\nl=0+2}{\thestep}{%increments \na in step of 2 5 times
    \uput{.5em}[270](\nl,0){\vphantom{$d_j$}\footnotesize \nl}%
  }%
  \multido{\i=0+1}{#1}{%
    \setcounter{proc}{\i+1}%
    \setlength{\ytmpa}{\i\psyunit+.14\psyunit}%
    % \rput(-0.5,\ytmpa){\small $\pi_{\theproc}$}%
    \setlength{\ytmpa}{\i\psyunit}%
    \setlength{\ytmpb}{\i\psyunit-.07\psyunit}%
    \setlength{\xtmpa}{#2\psxunit-0.5\psxunit}%
    \psline[linewidth=0.5pt]{->}(0,\ytmpa)(\xtmpa,\ytmpa)%
    \multido{\nt=0+1}{#2}{
      \psline[linestyle=solid,linewidth=1pt](\nt,\ytmpa)(\nt,\ytmpb)%
    }%
    \setcounter{step}{(#2-1)*4}%
    \multido{\nt=0+.25}{\thestep}{
      \psline[linestyle=solid,linewidth=0.5pt](\nt,\ytmpa)(\nt,\ytmpb)%
    }%
  }%
}%
\def\jobLocExec#1#2(#3,#4,#5){%
  \setcounter{proc}{#5-1}%
  \setlength{\xtmpa}{#3\psxunit}%
  \setlength{\ytmpa}{\theproc\psyunit}%
  \setlength{\xtmpb}{#3\psxunit+#4\psxunit}%
  \setlength{\ytmpb}{\theproc\psyunit+\execWidth}%
  \psset{linecolor=black,linestyle=solid}%
  \psframe[linecolor=black,linestyle=solid,fillstyle=none]
  (\xtmpa,\ytmpa)(\xtmpb,\ytmpb)%
  \ifthenelse{\equal{#1}{}}{}{%
  }{%
    \setlength{\xtmpa}{#3\psxunit+{#4\psxunit*\real{0.5}}}%
    \setlength{\ytmpb}{\theproc\psyunit+0.5\execWidth}%
    \ifthenelse{\equal{#2}{}}{%
      \renewcommand{\subscr}{$\mspace{3mu}$#1}%
    }{%
      \renewcommand{\subscr}{$\mspace{0.1mu}$#1$^{#2}$}%
    }%
    \rput(\xtmpa,\ytmpb){\scriptsize \subscr}%
  }%
}%
% #1: #task, #2: abs deadline, #3: #processor, #4: color
\def\jobdeadline#1(#2,#3){%
  \setcounter{proc}{#3-1}%
  \setlength{\xtmpa}{#2\psxunit}%
  \setlength{\ytmpa}{\theproc\psyunit}%
  \setlength{\ytmpb}{\theproc\psyunit+\execWidth+0.35\psyunit}%
  \psline[linestyle=solid,linewidth=0.5pt,arrows=-*,arrowsize=4pt]
  (\xtmpa,\ytmpa)(\xtmpa,\ytmpb)%
  \ifthenelse{\equal{#1}{}}{}{%
    \setlength{\ytmpa}{\theproc\psyunit+1.74\psyunit}%
    \uput{.3em}[270](\xtmpa,\ytmpa){\vphantom{$d_j$}\footnotesize #1}}%
}%

\psset{xunit=.68cm, yunit=.64cm}%
\begin{pspicture*}(-.5,-1.)(12.7,7)%
  \setlength{\ytrans}{2.7\psyunit}%
  \setlength{\ytmpa}{\ytrans-0.1\psyunit}%
  \setlength{\ytmpb}{\ytrans+3.7\psyunit}%
  \uput{.01em}[90](.3,\ytmpb){\vphantom{$d_j$}\footnotesize $C_s(t)$}%
  \setlength{\ytmpb}{\ytrans+3.6\psyunit}%
  \psline[linestyle=solid,linewidth=0.5pt,arrows=->,arrowsize=4pt]%
  (0,\ytmpa)(0,\ytmpb)%
  \setlength{\ytmpb}{\ytrans+3.6\psyunit}%
  \psline[linestyle=solid,linewidth=0.5pt,arrows=-<,arrowsize=4pt]%
  (3,\ytmpa)(3,\ytmpb)%
  \psline[linestyle=solid,linewidth=0.5pt,arrows=-<,arrowsize=4pt]%
  (6,\ytmpa)(6,\ytmpb)%
  \psline[linestyle=solid,linewidth=0.5pt,arrows=-<,arrowsize=4pt]%
  (8,\ytmpa)(8,\ytmpb)%
  \psline[linestyle=solid,linewidth=0.5pt,arrows=-<,arrowsize=4pt]%
  (12,\ytmpa)(12,\ytmpb)%
  \psline[linestyle=solid,linewidth=0.5pt,arrows=->](-0.1,\ytrans)(12.5,\ytrans)%
  \uput{.5em}[270](0,\ytrans){\vphantom{$d_j$}\footnotesize $0$}%
  \uput{.5em}[270](2.25,\ytrans){\vphantom{$d_j$}\footnotesize $\frac{9}{4}$}%
  \uput{.5em}[270](3,\ytrans){\vphantom{$d_j$}\footnotesize $3$}%
  \uput{.5em}[270](5.25,\ytrans){\vphantom{$d_j$}\footnotesize
    $3\!+\mspace{-5mu}\frac{9}{4}\;$}%
  \uput{.5em}[270](6,\ytrans){\vphantom{$d_j$}\footnotesize $6$}%
  \uput{.5em}[270](7.4,\ytrans){\vphantom{$d_j$}\footnotesize
    $6\!+\mspace{-5mu}\frac{3}{2}\;$}%
  \uput{.5em}[270](8.1,\ytrans){\vphantom{$d_j$}\footnotesize $8$}%
  \uput{.5em}[270](11,\ytrans){\vphantom{$d_j$}\footnotesize $11$}%
  \uput{.5em}[270](12,\ytrans){\vphantom{$d_j$}\footnotesize $12$}%
  % \uput{.5em}[270](12.5,\ytrans){\vphantom{$d_j$}\footnotesize $t$}%
  % 
  \setlength{\ytmpa}{\ytrans+2.25\psyunit}%
  \uput{.3em}[180](0,\ytmpa){\vphantom{$d_j$}\footnotesize $\frac{9}{4}$}%
  \setlength{\ytmpb}{\ytrans+1.5\psyunit}%
  \uput{.3em}[180](0,\ytmpb){\vphantom{$d_j$}\footnotesize $\frac{3}{2}$}%
  \setlength{\ytmpc}{\ytrans+3\psyunit}%
  \uput{.3em}[180](0,\ytmpc){\vphantom{$d_j$}\footnotesize $3$}%
  \psline[linestyle=solid,linewidth=1pt](0.0,\ytrans)(0.,\ytmpa)%
  \psline[linestyle=solid,linewidth=1pt](0.,\ytmpa)(2.25,\ytrans)%
  \psline[linestyle=solid,linewidth=1pt](2.25,\ytrans)(3,\ytrans)%
  \psline[linestyle=dashed,linewidth=0.5pt](0,\ytmpb)(6,\ytmpb)%
  \psline[linestyle=solid,linewidth=1pt](3,\ytrans)(3,\ytmpa)%
  \psline[linestyle=solid,linewidth=1pt](3,\ytmpa)(5.25,\ytrans)%
  \psline[linestyle=solid,linewidth=1pt](5.25,\ytrans)(6,\ytrans)%
  \psline[linestyle=solid,linewidth=1pt](6,\ytrans)(6,\ytmpb)%
  \psline[linestyle=solid,linewidth=1pt](6,\ytmpb)(7.5,\ytrans)%
  \psline[linestyle=solid,linewidth=1pt](7.5,\ytrans)(8,\ytrans)%
  \psline[linestyle=solid,linewidth=1pt](8.0,\ytrans)(8.,\ytmpc)%
  \psline[linestyle=solid,linewidth=1pt](8.,\ytmpc)(11,\ytrans)%
  \psline[linestyle=solid,linewidth=1pt](11,\ytrans)(12,\ytrans)%
  \psline[linestyle=dashed,linewidth=0.5pt](0,\ytmpc)(8,\ytmpc)%
  \schedAxes(1,13)%
  % \uput{.5em}[270](3,0){\vphantom{$d_j$}\footnotesize $3$}%
  % \uput{.5em}[270](12.5,0){\vphantom{$d_j$}\footnotesize $t$}%
  \uput{.3em}[180](0,.3){\vphantom{$d_j$}\footnotesize $\Sigma$}%
  % Task 1
  \jobLocExec{$1_1$}(0, 1, 1)%
  % \jobexec{1}(6, 2, 1)%
  \jobLocExec{$2_1$}(1, 1, 1)%
  \jobLocExec{$\mspace{-2mu}3$}(2, 0.25, 1)%
  \jobLocExec{$3_1$}(3, 0.75, 1)%
  \jobLocExec{$1_2$}(3.75, 1, 1)%
  \jobLocExec{$2_2$}(4.75, 0.5, 1)%
  \jobLocExec{$2_2$}(6, 0.5, 1)%
  \jobLocExec{$1_3$}(6.5, 1, 1)%
  \jobLocExec{$3_2$}(8, 1, 1)%
  \jobLocExec{$1_4$}(9, 1, 1)%
  \jobLocExec{$2_3$}(10, 1, 1)%
  \jobdeadline{$1_1$}(3,1)%
  \jobdeadline{$2_1$}(4,1)%
  \jobdeadline{$1_2, 3_1$}(6,1)%
  \jobdeadline{$2_2$}(8,1)%
  \jobdeadline{$1_3$}(9,1)%
  \jobdeadline{$1_4, 2_3, 3_2\quad$}(12,1)%
\end{pspicture*}%
\caption{Budget management and schedule of an EDF server $S$ with $\servSet(S) =
  \{ \taskdef(1,1/3, 3\Nat^* ), \taskdef(2,1/4, 4\Nat^* ), \taskdef(3,1/6,
  6\Nat^* ) \}$ and $\util(S) = 3/4$.}
\label{fig:budget}
\end{figure}

%\newpage
%obs-old: acho que vc queria dizer isto:
\begin{definition}\label{dfn:unitSet}
  A set $\servSet$ of fixed-utilization tasks is a unit set if
  $\;\util(\servSet) = 1$. The server $\serv(\servSet)$ associated to a unit set
  $\servSet$ is a unit server.
\end{definition}

In order to prove that EDF servers are predictable, we first present some
intermediate results.

\begin{definition}\label{dfn:scaling}
  Let $S$ be a server, $\servSet$ a set of servers with $\util(\servSet) \leq
  1$, and $\alpha$ a real such that $0 < \alpha \leq 1 / \util(S)$. The
  $\alpha$-scaled server of $S$ is the server with utilization $\alpha \util(S)$
  and deadlines equal to those of $S$. The $\alpha$-scaled set of $\servSet$ is
  the set of the $\alpha$-scaled servers of server in $\servSet$.
\end{definition}

As illustration, consider $\servSet= \{ S_1, S_2, S_3 \}$ a set of servers with
$\util(\servSet) = 0.5$, $\util(S_1) = 0.1$, $\util(S_2) = 0.15$ and $\util(S_3)
= 0.25$. The $2$-scaled set of $\servSet$ is $\servSet' = \{ S_1', S_2',
S_3' \}$ with $\util(\servSet') = 1$, $\util(S_1') = 0.2$, $\util(S_2') = 0.3$
and $\util(S_3') = 0.5$.
  
\begin{lemma}\label{lem:scalingEquiv}
  Let $\servSet$ be a set of EDF servers with $\util(\servSet) \leq 1$ and
  $\servSet'$ be its $\alpha$-scaled set. Define $S$ and $S'$ as two EDF servers
  associated to $\servSet$ and $\servSet'$ and consider that $\,\sched$ and
  $\sched'$ are their corresponding schedules, respectively. The schedule
  $\sched$ is feasible if and only if $\sched'$ is feasible.
\end{lemma}
  
\begin{IEEEproof}
  Suppose $\sched$ feasible. Consider a deadline $d$ in $\Lambda(S)$. Since $S$
  and $S'$ use EDF and $\Lambda(S) = \Lambda(S')$, $S$ and $S'$ execute their
  client jobs in the same order. As a consequence, all the executions of servers
  in $\clientOf(S)$ during $[0,d)$ must have a corresponding execution of a
  server in $\clientOf(S')$ during $[0,d)$.

  Also, since $S$ executes for $\util(S) d$ during $[0,d)$ and $\alpha \leq
  1/\util(S)$, the execution time $\util(S') d$ of $S'$ during $[0,d)$ satisfies
  $\alpha \util(S) d \leq d$.  Hence, a client job of $S'$ corresponding to an
  execution which completes in $\sched$ before $d$, completes before $d$ in
  $\sched'$. Since $\sched$ is feasible, this shows that $\sched'$ is feasible.

  To show that $\sched$ is feasible if $\sched'$ is feasible the same reasoning
  can be made with a scale equal to $\alpha' = 1/\alpha$
\end{IEEEproof}

%\newpage
\begin{lemma}\label{lem:edfServSched}
  The schedule of a set of servers $\servSet$ produced by the EDF server $S =
  \serv(\servSet)$ is feasible if and only if $\util(\servSet) \leq 1$.
\end{lemma}

\begin{IEEEproof}
  The proof presented here is an adaptation of the proof of Theorem $7$ from
  \cite{Liu73}. The difference between servers and tasks makes this presentation
  necessary.
  % Restringir este lema a utilização
  % 100\% do processador. Assim, não precisaríamos nos preocupar com idle
  % time. Como o teorema seguinte vc usa o fator de escala, vc acaba precisando
  % apenas de cenários onde a utilização do processador é igual a 1.

  First, assume that $\util(\servSet) > 1$. Let $[d,d')$ be a time interval with
  no processor idle time, where $d$ and $d'$ are two deadlines of servers in
  $\servSet$. By the assumed utilization, this time interval must exist. As the
  cumulated execution requirement within this interval is $\util(\servSet)(d'-d)
  > d'-d$, a deadline miss must occur, which shows the necessary condition.

  Suppose now that $d$ is the first deadline miss after time $t=0$ and let $S$
  be the server whose job $J$ misses its deadline at $d$.  Let $t'$ be the start
  time of the latest idle time interval before $d$. Assume that $t' = 0$ if such
  a time does not exist. Also, let $d'$ be the earliest deadline in $\Lambda(S)$
  after $t'$. Note that $d' < d$ otherwise no job would be released between
  $t'$ and $d$.
  % Let $t'$ be the latest idle time before $d$ (assume $t' = 0$ if such idle
  % time does not exist) and let $d'$ be the earliest deadline in $\Lambda(S)$
  % after $t'$. Acho que a definição de t'estava errada.
  If $d'$ is not equal to zero, then the processor must be idle just before
  $d'$.  Indeed, if there were some job executing just before $d'$, it would be
  released after $t'$ and its release instant would be a deadline in
  $\Lambda(S)$ occurring before $d'$ and after $t'$, which would contradict the
  definition of $d'$. Hence, only the time interval between $d'$ and $d$ is to
  be considered.  There are two cases to be distinguished depending on whether
  some lower priority server executes within $[d',d)$.
  % Indeed, suppose that there exist a job executing just before $d'$. Such a
  % job must have been released after $t'$ and its release instant is a deadline
  % in $\Lambda(S)$, before $d'$ and after $t'$, contradicting the minimality of
  % $d'$.

  \begin{figure}[h]
    \centering%
    \psset{xunit=1cm, yunit=1cm}%
    \begin{pspicture*}(0,-.6)(8,0.5)%
      \psline[linestyle=solid,linewidth=0.5pt,arrows=->](0,0)(8,0)%
      \multido{\nl=3.1+.1}{44}{%increments \na in step of 2 5 times
        \rput(\nl,0){%
          \psline[linestyle=solid,linewidth=0.5pt](-.1,-.1)(.1,.1)%
        } }%
      \multido{\nl=0.1+.1}{9}{%increments \na in step of 2 5 times
        \rput(\nl,0){%
          \psline[linestyle=solid,linewidth=0.5pt](-.1,-.1)(.1,.1)%
        } }%
      \uput{.5em}[270](2,0.7){ idle time}%
      \uput{.5em}[270](1,-.1){$t'$}%
      \psline[linestyle=solid,linewidth=1pt](1,-.14)(1,.14)%
      \uput{.5em}[270](3,-.1){$d'$}%
      \psline[linestyle=solid,linewidth=1pt](3.5,-.14)(3.5,.14)%
      \uput{.5em}[270](3.5,-.1){$d'_k$}%
      \psline[linestyle=solid,linewidth=1pt](3,-.14)(3,.14)%
      \uput{.5em}[270](6.5,-.1){$d_k$}%
      \psline[linestyle=solid,linewidth=1pt](6.5,-.14)(6.5,.14)%
      \uput{.5em}[270](7,-.1){$d$}%
      \psline[linestyle=solid,linewidth=1pt](7,-.14)(7,.14)%
    \end{pspicture*}%
    \caption{A deadline miss occurs for job $J$ at time $d$ and no job with
      lower priority than $J$ executes before $d$}
    \label{fig:deadlineMissA}
  \end{figure}
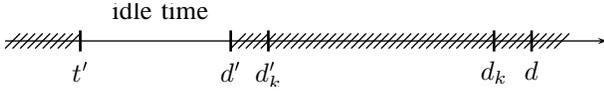

  \paragraph{Case 1} Illustrated by Figure \ref{fig:deadlineMissA}. Assume that
  no job of servers in $\clientOf(S)$ with lower priority than $J$ executes
  within $[d',d)$. Since there is no processor idle time between $d'$ and $d$
  and a deadline miss occurs at time $d$, it must be that the cumulated
  execution time of all jobs in $\clientOf(S)$ released at or after $d'$ and
  with deadline less than or equal to $d$ is strictly greater than
  $d-d'$. Consider servers $S_k$ whose jobs have their release instants and
  deadlines within $(d',d]$. Let $d_k'$ and $d_k$ be the first release instant
  and the last deadline of such jobs, respectively.  The cumulated execution
  time of such servers during $[d',d)$ equals $C = \sum_{S_k \in \clientOf(S)}
  \util(S_k)(d_k - d_k')$.  As $\sum_{S_k \in \clientOf(S)} \util(S_k) \leq
  \util(S) \leq 1$, $C \leq \util(S)(d - d') \leq d - d'$, leading to a
  contradiction.
 
  \begin{figure}[h]
    \centering%
    \psset{xunit=1cm, yunit=1cm}%
    \begin{pspicture*}(0,-.6)(8,0.5)%
      \psline[linestyle=solid,linewidth=0.5pt,arrows=->](0,0)(8,0)%
      \multido{\nl=3.1+.1}{44}{%increments \na in step of 2 5 times
        \rput(\nl,0){%
          \psline[linestyle=solid,linewidth=0.5pt](-.1,-.1)(.1,.1)%
        } }%
      \multido{\nl=0.1+.1}{9}{%increments \na in step of 2 5 times
        \rput(\nl,0){%
          \psline[linestyle=solid,linewidth=0.5pt](-.1,-.1)(.1,.1)%
        } }%
      \uput{.5em}[270](2,0.7){ idle time}%
      \uput{.5em}[270](1,-.1){$t'$}%
      \psline[linestyle=solid,linewidth=1pt](1,-.14)(1,.14)%
      \uput{.5em}[270](3,-.1){$d'$}%
      \psline[linestyle=solid,linewidth=1pt](4,-.14)(4,.14)%
      \uput{.5em}[270](4,-.1){$d''$}%
      \psline[linestyle=solid,linewidth=1pt](3,-.14)(3,.14)%
      \uput{.5em}[270](5.5,-.1){\vphantom{$d_j$}$r$}%
      \psline[linestyle=solid,linewidth=1pt](5.5,-.14)(5.5,.14)%
      \uput{.5em}[270](7,-.1){$d$}%
      \psline[linestyle=solid,linewidth=1pt](7,-.14)(7,.14)%
    \end{pspicture*}%
    \caption{ A deadline miss occurs for job $J$ at time $d$ and some lower
      priority job than $J$ executes before $d$}
    \label{fig:deadlineMissB}
  \end{figure}
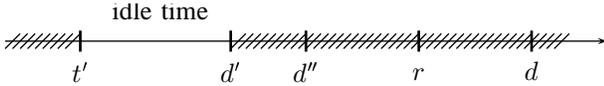
  
  \paragraph{Case 2} Illustrated by Figure \ref{fig:deadlineMissB}. Assume that
  there exist client jobs of $S$ with lower priority than $J$ that execute
  within $[d',d)$. Let $d''$ be the latest deadline after which no such jobs
  execute and consider $r$ the release instant of $J$. Since $J$ misses its
  deadline, no job with lower priority than $J$ can execute after $r$. Thus, we
  must have $d'' \leqslant r < d$. Also, there is no processor idle time in
  $[d'',d)$. Thus, for a deadline miss to occur at time $d$, it must be that the
  cumulated execution time of all servers in $\clientOf(S)$ during $[d'',d)$ is
  greater than $d-d''$.

  Also, it must be that a lower priority job was executing just before $d''$.
  Indeed, if $J'$, a job with higher priority than $J$, was executing just
  before $d''$, its release time $r'$ would be before $d''$ and no job with
  lower priority than $J$ could have executed after $r'$, contradicting the
  minimality of $d''$. Thus, no job released before $d''$ and with higher
  priority than $J$ executes between $d''$ and $d$.  Hence, the jobs that
  contribute to the cumulated execution time during $[d'',d)$ must have higher
  priorities than $J$ and must be released after $d''$. The cumulated
  requirement of such jobs of a server $S_k$ is not greater than $\util(S_k)
  (d-d'')$. Henceforth, since $\sum_{S_k \in \clientOf(S)} \util(S_k) = \util(S)
  \leq 1$, the cumulated execution time of all servers during $[d'',d)$ cannot
  be greater than $\util(S)(d-d'') \leq d-d''$, reaching a contradiction.
\end{IEEEproof}

\begin{theorem}\label{thm:edfServPredict}
  An EDF server is predictable.
\end{theorem}

\begin{IEEEproof}
  Consider a set of servers $\servSet = \{S_1, S_2, \ldots, S_n\}$ such that
  $\util(\servSet) \leq 1$ and assume that $\servSet$ is to be scheduled by an
  EDF server $S$. Let $\servSet'$ be the $1/\util(\servSet)$-scaled server set
  of $\servSet$. Hence, by Definition \ref{dfn:scaling}, we have
  $\util(\servSet') = \sum_{i=1}^n \util(S_i) / \util(\servSet) = 1$. Let $S'$
  be the EDF server associated to $\servSet'$.  By Lemma \ref{lem:scalingEquiv},
  the schedule $\sched$ of $\servSet$ by $S$ is feasible if and only if the
  schedule $\sched'$ of $\servSet'$ be $S'$ is feasible.  But, $S'$ schedules
  servers as EDF. Indeed, consider a release instant $r$ of $S'$ at which the
  budget of $S'$ is set to $\lambda_{S'}(r) - r$. During the entire interval
  $[r, \lambda_{S'}(r))$, the budget of $S'$ is strictly positive. This implies
  that $S'$ is not constrained by its budget during the whole interval $[r,
  \lambda_{S'}(r))$. Thus, $S'$ behaves as if it has infinite budget and
  schedules its client servers according to EDF. Since, by Lemma
  \ref{lem:edfServSched}, a server set of utilization one is feasible by EDF,
  the schedule $\sched'$ produced by $S'$ is feasible and so is $\sched$.
\end{IEEEproof}
% obs-olhar: Paul, note que modifiquei um pouco o início da prova. O conjunto
% servSet pode conter qualquer outro servidor que possivelmente bloqueia algum
% S_i. Então, o lema que você enunciou em seguida não é necessário. Coloquei o
% parágrafo abaixo para fazer uma conexão com a seção seguinte.

It is worth saying that Theorem \ref{thm:edfServPredict} implicitly assumes that
server $S$ executes on possibly more than one processor. The client servers of
$S$ do not execute in parallel, though. The assignment of servers to processors
is carried out on-line and is specified in the next section.

\section{Virtual Scheduling}\label{sec:virtualSched}

In this section we present two basic operations, dual and packing, which are
used to transform a multiprocessor system into an equivalent uniprocessor
system. The schedule for the found uniprocessor system is produced on-line by
EDF and the corresponding schedule for the original multiprocessor system is
deduced straightforwardly by following simple rules. The transformation
procedure can generate one or more virtual systems, each of which with fewer
processors than the original (real) system.

The dual operation, detailed in Section \ref{sec:dualSched}, transforms a
fixed-utilization task $\tau$ into another task $\tau^*$ representing the slack
task of $\tau$ and called the dual task of $\tau$. That is $\util(\tau^*) = 1 -
\util(\tau)$ and the deadlines of $\tau^*$ are equal to those of $\tau$. As
$\util(\tau) > 0.5$ implies $\util(\tau^*) < 0.5$, the dual operation plays the
role of reducing the utilization of the system made of complementary dual tasks
as compared to the original system.

The packing operation, presented in Section \ref{sec:packing}, groups one or
more tasks into a server. As fixed-utilization tasks whose utilization do not
sum up more than $1$ can be packed into a single server, the role of the packing
operation is to reduce the number of tasks to be scheduled.

By performing a pair of dual and packing operations, one is able to create a
virtual system with less processor and tasks. Hence, it is useful to have both
operations composed into a single one, called reduction operation, which will be
defined in Section \ref{sec:dualPackingSched}. As will be seen in Section
\ref{sec:reducCorrectness}, after performing a series of reduction operation,
the schedule of the multiprocessor system can be deduced from the (virtual)
schedule of the transformed uniprocessor system. Although a reduction from the
original system into the virtual ones is carried out off-line, the generation of
the multiprocessor schedule for the original system can be done on-line.
Section IV.E ilustrates the proposed approach with an example.

\subsection{Dual Operation}\label{sec:dualSched}

As servers are actually fixed-utilization tasks and will be used as a basic
scheduling mechanism, the dual operation is defined for servers.

\begin{definition}[Dual Server]\label{dfn:dualServer}
  Let $S$ be a server with utilization $\util(S)$ such that $0 < \util(S) <
  1$. The dual server of $S$ is defined as the server $S^*$ whose utilization
  $\util(S^*) = 1 - \util(S)$, deadlines are equal to those of $S$ and
  scheduling algorithm identical to that of $S$. If $\;\servSet$ is a set of
  servers, then the dual set $\servSet^*$ of $\servSet$ is the set of servers
  which are duals of the servers in $\servSet$, i.e. $S \in \servSet$ if and
  only if $S^* \in \servSet^*$.
\end{definition}

Note that servers with utilization equal to $1$ or $0$ are not considered in
Definition \ref{dfn:dualServer}. This is not a problem since in these cases $S$
can straightforwardly be scheduled. Indeed, if $S$ is a server with $100\%$
utilization, a processor can be allocated to $S$ and by Theorem
\ref{thm:edfServPredict}, all clients of $S$ meet their deadlines. In case that
$S$ is a null-utilization server, it is enough to ensure that $S$ never gets
executing.

We define the bijection $\dual$ from a set of non-integer (neither zero nor one)
utilization servers $\servSet$ to its dual set $\servSet^*$ as the function
which associates to a server $S$ its dual server $S^*$, i.e $\dual(S) = S^*$.

% Note that the utilization of a server is a static parameter,
% defined before the execution of the system. Thus, a dual server is well-defined,
% and, through the encapsulation of tasks in servers, we can apply the duality
% approach to servers in order to schedule task sets. However, there exist task
% sets for which the dual approach can not be applied directly. Indeed, if a
% server $S$ has an utilization equal to one, then its dual server $S^*$ has a
% null utilization.

\begin{definition}[Dual Schedule]\label{dfn:dualSchedule}
  Let $\servSet$ be a set of servers and $\servSet^*$ be its dual set.  Two
  schedules $\sched$ of $\servSet$ and $\sched^*$ of $\servSet^*$ are duals if,
  at any time, a server $S$ in $\servSet$ executes in $\sched$ if and only if
  its dual server $S^*$ does not execute in $\sched^*$.
\end{definition}
%It is clear that for any schedule $\sched$, $(\sched^*)^*= \sched$.

The following theorem relates the feasibility of a set of servers to the
feasibility of its dual set. It is enunciated assuming a fully utilized
system. However, recall from Section \ref{sec:fullyUtilSyst} that any system can
be extended to a fully utilized system in order to apply the results presented
here.

\begin{theorem}[Dual Operation]\label{thm:dualSched}
  Let $\;\servSet = \{ S_1, S_2, \ldots, S_n \}$ be a set of $n = m + k$ servers
  with $k \geqslant 1$ and $\util(\servSet) = m$. The schedule $\sched$ of
  $\servSet$ on $m$ processors is feasible if and only if its dual schedule
  $\sched^*$ is feasible on $k$ processors.
\end{theorem}

\begin{IEEEproof}
  In order to prove the necessary condition, assume that a schedule of
  $\servSet$ on $m$ processors, $\sched$, is feasible.  By Definition
  \ref{dfn:dualSchedule}, we know that $S_i$ executes in $\sched$ whenever
  $S_i^*$ does not execute in $\sched^*$, and vice-versa. Now, consider the
  executions in $\sched \times \sched^*$ of a pair $(S_i,S_i^*)$ and define a
  schedule $\bar{\sched}$ for the set $\bar{\servSet} = \servSet \cup
  \servSet^*$ as follows: $S_i$ always executes on the same processor in
  $\bar{\sched}$; $S_i$ executes in $\bar{\sched}$ at time $t$ if and only if it
  executes at time $t$ in $\sched$; and whenever $S_i$ is not executing in
  $\sched$, $S_i^*$ is executing in $\bar{\sched}$ on the same processor as
  $S_i$.
  
  By construction, the executions of $\servSet$ and $\servSet^*$ in
  $\bar{\sched}$ correspond to their executions in $\sched$ and $\sched^*$,
  respectively. Also, in $\bar{\sched}$, $S_i$ and $S_i^*$ execute on a single
  processor. Since $\util(S_i) + \util(S_i^*) = 1$ and $S_i$ and $S_i^*$ have
  the same deadlines, the feasibility of $S_i$ implies the feasibility of
  $S_i^*$. Since this is true for all pairs $(S_i, S_i^*)$, we deduce that both
  $\sched^*$ and $\bar{\sched}$ are feasible. Furthermore, as by the definition
  of $\bar{\sched}$, $n=m+k$ processors are needed and by assumption $\sched$
  uses $m$ processors, $\sched^*$ can be constructed on $k$ processors.

  The proof of the sufficient condition is symmetric and can be shown using
  similar arguments.
\end{IEEEproof}

Theorem \ref{thm:dualSched} does not establish any scheduling rule to generate
feasible schedules. It only states that determining a feasible schedule for a
given server set on $m$ processors is equivalent to finding a feasible schedule
for the transformed set on $n-m$ virtual processors.  Nonetheless, this theorem
raises an interesting issue. Indeed, dealing with $n-m$ virtual processors
instead of $m$ can be advantageous if $n-m < m$. In order to illustrate this
observation, consider a set of three servers with utilization equal to
$2/3$. Instead of searching for a feasible schedule on two processors, one can
focus on the schedule of the dual servers on just one virtual processor, a
problem whose solution is well known. In order to guarantee that dealing with
dual servers is advantageous, the packing operation plays a central role.

%\vspace{0.cm}
\subsection{Packing Operation}
\label{sec:packing}

As seen in the previous section, the dual operation is a powerful mechanism to
reduce the number of processors but only works properly if $n-m < m$. If this is
not the case, one needs to reduce the number of servers to be scheduled,
aggregating them into servers. This is achieved by the packing operation, which
is formally described in this section.

%\vspace{0.0cm}
\begin{definition}[Packed Server Set]\label{dfn:packedSet}
  A set of non-zero utilization servers $\servSet$ is packed if it is a
  singleton or if $|\servSet| \geq 2$ and for any two distinct servers $S$ and
  $S'$ in $\servSet$, $\util(S) + \util(S') > 1$.
\end{definition}

% Given a set of servers $\servSet$, we define $\unitPower(\servSet)$ as the set
% of all subsets of $\servSet$ whose utilization is less than or equal to one.

% % obs-paul: def k ???
% \begin{definition}[Packing Operation]\label{dfn:packMap}
%   Let $\servSet$ be a set of non-zero utilization servers. A packing operation
%   $\pack$ on $\servSet$ is a map from $\servSet$ to $\unitPower(\servSet)$
%   which splits $\servSet$ into a partition of $k$ disjunct subsets
%   $(\servSet_i)_{1 \leq i \leqslant k }$ such that $\{ \serv(\servSet_i), i = 1,
%   \ldots, k \}$ is a packed server set.
% \end{definition}

% \vspace{0.0cm}
% \begin{definition}[Packing Operation]\label{dfn:packMap}
%   Let $\servSet$ be a set of non-zero utilization servers. A packing operation
%   $\pack$ on $\servSet$ defines a partition on $\servSet$, denoted
%   $\pack(\servSet)$, such that $\serv(\pack(\servSet))$ is a packed server set.
% \end{definition}

%\vspace{0.0cm}

\begin{definition}[Packing Operation]\label{dfn:packMap}
  Let $\servSet$ be a set of non-zero utilization servers. A packing operation
  $\pack$ associates a packed set of servers $\pack(\servSet)$ to $\servSet$
  such that the set collection $(\clientOf(S))_{S \in \pack(\servSet)}$ is a
  partition of $\servSet$.
\end{definition}

% By this definition, if $k = |\pack(\servSet)|$ and $\pack(\servSet) =
% (\servSet_i)_{1 \leq i \leqslant k }$, then $\serv(\pack(\servSet)) = \{
% \serv(\servSet_i), i = 1, \ldots, k \}$ with $\clientOf(S_i) \in
% \unitPower(\servSet)$.

% the image by $\pack$ of a server $S_j$ in $\servSet$ is the set of servers
% $\servSet_i$ such that $S_j$ is in $\servSet_i$.  Hence, if $\pack(S_j) =
% \servSet_i$ then $S_j \in \servSet_i$.

%\vspace{0.cm}
Note that a packing operation is a projection ($\pack \comp \pack = \pack$)
since the packing of a packed set is the packed set itself.

%Since $(\servSet_i)_{1 \leqslant i \leqslant k}$ is a partition of $\servSet$,
% A packing operation is an equivalence relation on $\servSet$ whose equivalence
% classes are the servers $\servSet_i$ for $i$ in $\{1, \ldots, k\}$.

An example of partition, produced by applying a packing operation on a set
$\servSet$ of $10$ servers, is illustrated by the set $\pack(\futSet)$ on the
top of Figure \ref{fig:dualSet}. In this example, the partition of $\servSet$ is
comprised of the three sets $\clientOf(S_{11})$, $\clientOf(S_{12})$ and
$\clientOf(S_{13})$. As an illustration of Definition \ref{dfn:packMap}, we
have, $\clientOf(S_{11}) = \pack(S_2) = \pack(S_3) = \pack(S_7)$.

%\vspace{0.cm}
\begin{lemma}\label{lem:packMap}
  Let $\servSet$ be a set of non-zero utilization servers. If $\,\pack$ is a
  packing operation on $\servSet$, then $\util(\pack(\servSet)) =
  \util(\servSet)$ and $|\pack(\servSet)| \geqslant \util(\servSet)$.
\end{lemma}

\begin{IEEEproof}
  A packing operation does not change the utilization of servers in $\servSet$
  and so $\util(\pack(\servSet)) = \util(\servSet)$. To show the inequality,
  suppose that $\util(\servSet) = k + \vareps$ with $k$ natural and $0 \leqslant
  \vareps < 1$. As the utilization of a server is not greater than one, there
  must exist at least $\lceil k+ \vareps\rceil$ servers in $\pack(\servSet)$.
\end{IEEEproof}

%\newpage
The following lemma establishes an upper bound on the number of servers resulted
from packing an arbitrary number of non-zero utilization servers with total
utilization $u$.

\begin{lemma}\label{lem:packReduc}
  If $\servSet$ is a set of non-zero utilization servers and $\servSet$ is
  packed, then $|\servSet| < 2 \util(\servSet)$.
\end{lemma}

\begin{IEEEproof}
  Let $ n = |\servSet|$ and $u_i = \mu(S_i)$ for $S_i \in \servSet$.  Since
  $\servSet$ is packed, there exists at most one server in $\servSet$, say
  $S_n$, such that $u_n < 1/2$. All other servers have utilization greater that
  $1/2$.  Thus, $\sum_{i=1}^{n-2} u_i > (n-2)/2$. As $u_{n-1} + u_n > 1$, it
  follows that $\sum_{i=1}^n u_i = \util(\servSet) > n/2$.
\end{IEEEproof}

\subsection{Reduction Operation}
\label{sec:dualPackingSched}

In this section we define the composition of the dual and packing operations. We
begin by noting that the following relation holds.

\begin{lemma}\label{lem:dualUtil}
  If $\,\servSet$ is a packed server set with more than one server, then
  $\mu(\dual (\servSet)) < (|\servSet| + 1)/2$.
\end{lemma}
%obs-old: alterei a eq. do lema. antes o limite superior estava maior.

\begin{IEEEproof}
  As $\servSet$ is packed, at least $|\servSet| - 1$ servers have their
  utilization strictly greater than $1/2$. Thus, at least all but one server in
  $\dual(\servSet)$ have utilization strictly less than
  $1/2$. Hence, $\mu(\dual(\servSet)) < (|\servSet|-1)/2+1$.
\end{IEEEproof}

\begin{figure}[t]
  \centering%
  \psset{xunit=0.84cm, yunit=0.54cm}%
  \begin{pspicture*}(-1,-2)(8.1,9.4)%
    \psset{fillstyle=none,linewidth=.5pt}%
    % \psgrid[subgriddiv=1,gridwidth=.5pt,griddots=4,subgriddots=0,gridlabels=.3]
    % (-0.5,-1.5)(10,10)%
    \uput{0}[0](6.5,8.8){\vphantom{$d_j$} $\pack(\servSet^{(2)})$}%
    \psellipse[fillstyle=none](3.7,5.8)(4.35,3.5)%
    \psellipse[fillstyle=none](2.1,7.2)(1.4,1.4)%
    \uput{0}[0](3.,8.5){\vphantom{$d_j$} $S_{11}^{\;(0.4)}$}%
    \uput{0}[0](0.9,7.8){\vphantom{$d_j$} $S_2^{\;(0.2)}$}%
    \uput{0}[0](1.5,7.2){\vphantom{$d_j$} $S_3^{\;(0.1)}$}%
    \uput{0}[0](2.1,6.6){\vphantom{$d_j$} $S_7^{\;(0.1)}$}%
    \psellipse[fillstyle=none](5.2,5.1)(1.8,1.7)%
    \uput{0}[0](6.6,6.4){\vphantom{$d_j$} $S_{12}^{\;(0.9)}$}%
    \uput{0}[0](4.7,6){\vphantom{$d_j$} $S_1^{\;(0.4)}$}%
    \uput{0}[0](5.5,5.4){\vphantom{$d_j$} $S_4^{\;(0.2)}$}%
    \uput{0}[0](3.5,5.4){\vphantom{$d_j$} $S_5^{\;(0.1)}$}%
    \uput{0}[0](4.3,4.8){\vphantom{$d_j$} $S_8^{\;(0.1)}$}%
    \uput{0}[0](5.1,4.2){\vphantom{$d_j$} $S_{10}^{\;(0.1)}$}%
    \psellipse[fillstyle=none](1.6,4)(1.1,0.9)%
    \uput{0}[0](2.1,5.1){\vphantom{$d_j$} $S_{13}^{\;(0.7)}$}%
    \uput{0}[0](0.7,4.3){\vphantom{$d_j$} $S_6^{\;(0.4)}$}%
    \uput{0}[0](1.3,3.7){\vphantom{$d_j$} $S_9^{\;(0.3)}$}%
    \psbezier[linewidth=0.5pt,showpoints=false]{->}(1,6.)(0,4.5)(0,2.5)(0.6,0.5)
    \psbezier[linewidth=0.5pt,showpoints=false]{->}(2.5,3.2)(2.8,3)(3.4,1.5)(3.5,0.5)
    \psbezier[linewidth=0.5pt,showpoints=false]{->}(5.5,3.2)(5.7,2.5)(5.7,1.5)(5.5,0.5)
    % 
    % \uput{0}[0](5,2.5){\vphantom{$d_j$} $\reduc(\servSet) =
    % \dual(\serv(\pack(\servSet)))^{(1)}$}%
    \uput{0}[0](6.5,1.5){\vphantom{$d_j$} $\reduc(\servSet)^{(1)}$}%
    \psellipse[fillstyle=none](3.6,0)(4.3,1.4)%
    \uput{0}[0](0.,0){\vphantom{$d_j$} $\dual(S_{11})^{(0.6)}$}%
    \uput{0}[0](5,0){\vphantom{$d_j$} $\dual(S_{12})^{(0.1)}$}%
    \uput{0}[0](2.5,0){\vphantom{$d_j$} $\dual(S_{13})^{(0.3)}$}%
  \end{pspicture*}%
  \caption{Partition $\pack(\servSet)$ of $\servSet = \{ S_1, S_2, \ldots ,
    S_{10} \}$ into three subsets $\clientOf(S_{11}) = \pack(S_2)$,
    $\clientOf(S_{12}) = \pack(S_5)$ and $\clientOf(S_{13}) = \pack(S_6)$ and
    image $\reduc(\servSet)$ of $\servSet$. The utilization $u$ of a server $S$
    or a set of server $\servSet$ is indicated by the notation $S^{(u)}$ and
    $\servSet^{(u)}$, respectively.}
  \label{fig:dualSet}
\end{figure}
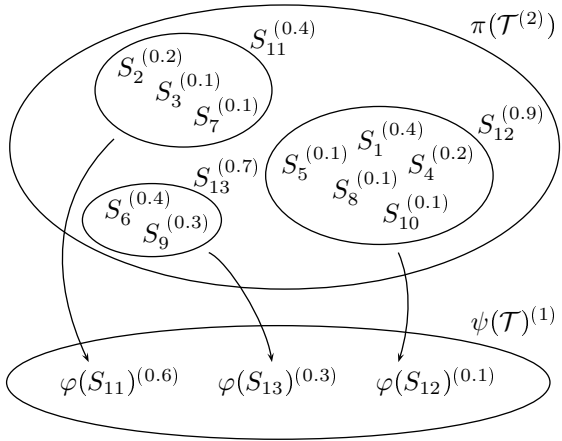

According to Lemma \ref{lem:dualUtil}, the action of the dual operation applied
to a packed set allows for the generation of a set of servers whose total
utilization is less than the utilization of the original packed set, as
illustrated in Figure \ref{fig:dualSet}. Considering an integer utilization
server set $\servSet$, this makes it possible to reduce the number of servers
progressively by carrying out the composition of a packing operation and the
dual operation until $\servSet$ is reduced to a set of unit servers. Since this
server can be scheduled on a single processor, as will be shown later on, it is
known by Theorem \ref{thm:dualSched} that a feasible schedule for the original
multiprocessor systems can be derived. Based on these observations it is worth
defining a reduction operation as the composition of a packing operation and the
dual operation.

\begin{definition}
  A reduction operation on a set of servers $\servSet$, denoted
  $\reduc(\servSet)$, is the composition of the dual operation $\dual$
  (Definition \ref{dfn:dualSchedule}), with a packing operation $\pack$
  (Definition \ref{dfn:packMap}), namely $\reduc = \dual \comp \pack$.
\end{definition}

% It is important to notice that a packing operation returns the servers a set of
% servers, and not their server, it is necessary to apply the operator $\serv$
% after a packing operation and before the dual operation. 
The action of the operator $\reduc$ on a set $\servSet$ of $10$ servers is
illustrated in Figure \ref{fig:dualSet}.

\subsection{Reduction Correctness}
\label{sec:reducCorrectness}

The results shown in the previous sections will be used here to show how to
transform a multiprocessor system into an equivalent (virtual) uniprocessor
system by carrying out a series of reduction operations on the target system.
First, it is shown in Lemma \ref{lem:reducConv} that a reduction operator
returns a reduced task system with smaller cardinality.
% First, it is shown in Lemma \ref{lem:reducConv} that a reduction operator
% returns a virtual system of smaller size.
Then, Lemma \ref{lem:threeSetInteger} and Theorem \ref{thm:reducConv} show that
after performing a series of reduction operations, a set of servers can be
transformed into a unit server, which, according to Theorem \ref{thm:reduction},
can be used to generate a feasible schedule on a uniprocessor system. Finally,
it is shown in Theorem \ref{thm:complexity} that time complexity for carrying
out the necessary series of reduction operations is dominated by the time
complexity of the packing operation.

\begin{lemma}\label{lem:reducConv}
  If $\servSet$ is a packed set of non-unit servers,
  \[
  | \pack \comp \dual(\servSet) | \leq \left \lceil \frac{|\servSet| + 1}{2}
  \right \rceil
  \]
\end{lemma}

\begin{IEEEproof}
  Let $n = |\servSet|$. By the definition of $\servSet$, which is packed, there
  is at most one server $S_i$ in $\servSet$ so that $\util(S_i) \leqslant
  1/2$. This implies that at least $n-1$ servers in $\dual(\servSet)$ have their
  utilizations less than $1/2$. Since servers in $\servSet$ are non-unit
  servers, their duals are non-zero-utilization servers. Hence, those dual
  servers can be packed up pairwisely, which implies that there will be at most
  $\lceil (n-1)/2 \rceil + 1$ servers after carrying out the packing operation.
  Thus, we deduce that $|\pack \comp \dual(\servSet)| \leq \lceil (n+1)/2
  \rceil$.
\end{IEEEproof}

\begin{lemma}\label{lem:threeSetInteger}
  Let $\servSet$ be a packed set of non-unit servers. If $\util(\servSet)$ is
  an integer, then $|\servSet| \geq 3$.
\end{lemma}

% \begin{IEEEproof}
%   Since $\servSet$ is a packed set of non-unit servers, by Definition
%   \ref{dfn:packMap}, there must exists at least two servers in $\servSet$, say
%   $S_1$ and $S_2$ with $0 < \util(S_1) < 1$ and $0 < \util(S_2) < 1$. Now,
%   suppose that $\servSet = \{ S_1, S_2 \}$. Since $\util(\servSet)$ is an
%   integer, $\util(S_1) + \util(S_2) = \util(\servSet)$ must equal one. However,
%   this is impossible since $\util(S_1) + \util(S_2) > 1$ by Definition
%   \ref{dfn:packMap} of a packed set. Also, $\util(S_1) + \util(S_2) < 2$. Thus,
%   there must exist at least another server in $\servSet$ since $\util(\servSet)$
%   is an integer, and so, we deduce that $|\pack(\servSet)| \geq 3$.
% \end{IEEEproof}

\begin{IEEEproof}
  If $|\servSet| \leq 2$, $\servSet$ would contain a unit server since
  $\util(\servSet)$ is a non-null integer. $\,$ Nonetheless, there exist
  larger non-unit server sets. For example, let $\servSet$ be a set of
  servers such that each server in $\servSet$ has utilization
  $\util(\servSet)/|\servSet|$ and $|\servSet| = \util(\servSet) + 1$.
\end{IEEEproof}

% \begin{lemma}\label{lem:threeSetReduc}
%   Let $\servSet = \{ S_1, S_2, S_3 \}$ be a packed set of three non-zero
%   utilization servers. If $\util(\servSet) = 2$ then $\dual(\servSet)$ is a
%   unit server set.
% \end{lemma}

% \begin{IEEEproof}
%   \[
%   \util(\dual(\servSet)) = \sum_{S \in \servSet} \util(\dual(S)) = \sum_{S \in
%   \servSet} (1 - \util(S)) = 3 - 2 = 1
%   \]
% \end{IEEEproof}

% obs-olhar: a def. de reducão ainda me parece incorreta. Fazendo i=1, o que
% seria a composição de \servset com \reduc(\servset)?

\begin{definition}[Reduction Level and Virtual Processor]\label{dfn:virtalProc}
  Let $i$ be a natural greater than one. The operator $\,\reduc^i$ is
  recursively defined as follows $\reduc^0(\servSet) = \servSet$ and
  $\reduc^{i}(\servSet) = \reduc \comp \reduc^{i-1}(\servSet)$. The server
  system $\reduc^i(\servSet)$ is said to be at reduction level $i$ and is to be
  executed on a set of virtual processors.
\end{definition}

Table \ref{tab:reducExample} illustrates a reduction of a system composed of
$10$ fixed-utilization tasks to be executed on $6$ processors. As can be
seen, two reduction levels were generated by the reduction operation. At
reduction level $1$, three virtual processors are necessary to schedule the $8$
remaining servers, while at reduction level $2$, a single virtual processor
suffices to schedule the $3$ remaining servers.

The next theorem states that the iteration of the operator $\reduc$ transforms a
set of servers of integer utilization into a set of unit servers. For a given
set of servers $\servSet$, the number of iterations necessary to achieve this
convergence to unit servers vary for each initial server in $\servSet$, as shown
in Table \ref{tab:reducExample}.

\setlength{\tabcolsep}{0.5em}
\begin{table}
  \centering
\caption{Reduction Example of a Set of Servers.\label{tab:reducExample}}
  \begin{tabular}{|l|c|c|c|c|c|c|c|c|c|c|}
    \hline  \rule{0cm}{.4cm}
    & \multicolumn{10}{c|}{Server Utilization \rule{0cm}{.4cm}} \\
    \hline 
    $\reduc^0(\servSet)$ \rule{0cm}{.35cm} & .6 & .6 & .6 & .6 & .6 & .8 & .6 & .6 & .5 & .5 \\
    \hline 
    $\pack(\reduc^0(\servSet))$ \rule{0cm}{.35cm} & .6 & .6 & .6 & .6 & .6 & .8 & .6 & .6 & 
    \multicolumn{2}{c|}{1} \\
    \hline  
    $\reduc^1(\servSet)$ \rule{0cm}{.35cm} & .4 & .4 & .4 & .4 & .4 & .2 & .4 & .4 \\
    \cline{1-9}
    $\pack (\reduc^1(\servSet))$ \rule{0cm}{.35cm} & \multicolumn{2}{c|}{.8} & 
    \multicolumn{2}{c|}{.8} & .4 & \multicolumn{3}{c|}{1} \\
    \cline{1-9}
    $\reduc^2(\servSet)$ \rule{0cm}{.35cm} & \multicolumn{2}{c|}{.2} & 
    \multicolumn{2}{c|}{.2} & .6 \\
    \cline{1-6} 
    $\pack (\reduc^2(\servSet))$ & \multicolumn{5}{c|}{ \rule{0cm}{.35cm} 1}\\
    \cline{1-6}
  \end{tabular}
\end{table}

% \newpage
\begin{theorem}[Reduction Convergence]\label{thm:reducConv}
  Let $\servSet$ be a set of non-zero utilization servers.  If $\,\servSet$ is a
  packed set of servers with integer utilization, then for any element $S \in
  \servSet$, $\pack(\reduc^p(S))$ is a unit server set for some level $p \geq
  1$.
\end{theorem}

%obs-paul: reler esta prova.
\begin{IEEEproof}
  $\pack(\reduc^0(\servSet))$ can be seen as a partition comprised of two
  subsets, those that contain unit sets and those that do not.  Let $F_{0}$ and
  $U_{0}$ be these sets, formally defined as follows: $F_0 = \{S \in
  \pack(\reduc^0(\servSet)), \util(S)) < 1\}$; $U_0 = \{S \in
  \pack(\reduc^0(\servSet)), \util(S)) = 1\}$; $F_0 \cup U_0 =
  \pack(\reduc^0(\servSet))$.  Also, for $k>0$ define $F_k$ and $U_k$ as $F_k =
  \{S \in \pack(\reduc(F_{k-1})), \util(S)) < 1\}$ and $U_k = \{S \in
  \pack(\reduc(F_{k-1})), \util(S)) = 1\}$.  We first claim that while $F_{k-1}
  \not= \{\}$, $|F_{k}| < |F_{k-1}|$ and that $\mu(F_k)$ is integer.  We show
  the claim by induction on $k$.

  \paragraph{Base case}
  As $F_0 \cup U_0$ is a packed set with integer utilization, it follows that
  $\mu(F_0)$ is also integer since $U_0$ is a unit set.  Consider $F_1$ and
  $U_1$ the partition of $\pack(\reduc^1(F_0))$.  As $F_1 \cup U_1 =
  \pack(\reduc^1(F_0))$ and $U_0$ have integer utilization, $\mu(F_1)$ is also
  integer.  Also, by Lemma \ref{lem:threeSetInteger}, $|F_1| \geq 3$ and $F_1$
  is a packed set of servers, we deduce from Lemma \ref{lem:reducConv} that
  \[
  3 \leq |F_{1}| \leq \left \lceil \frac{|F_0| + 1}{2} \right \rceil
  \]
  Therefore, $|F_1| < |F_0|$, since $\lceil (x+1)/2 \rceil < x$ for $x \geq 3$.

  \paragraph{Induction step}
  Assuming the claim holds until $k-1$, it can be shown that it holds for $k$
  analogously as it was done for the base case.
  \paragraph{Conclusion}
  By the claim there must exist $k$ such that $F_k = \{\}$ since by Lemma
  \ref{lem:threeSetInteger} there is no $F_k$ such that $|F_k| < 3$.  Hence,
  $\pack(\reduc^{p}(S))$ must belong to some $U_{p}$ for some $p\leq k$, which
  completes the proof.
\end{IEEEproof}
  
% obs-olhar: acho que devemos colocar um exemplo para deixar claro que proper
% set está associado com a operação de redução. Não sei se entendi direito esta
% definição, apesar de saber o que se quer dizer. Temos que checar sua
% consistência.

\begin{definition}[Proper Server Set]
  Let $\reduc = \dual \,\comp\, \pack$ be a reduction operation and
  $\servSet$ be a set of servers with $\util(\servSet) \in \Nat^*$. A subset of
  $\servSet$ is proper for $\reduc$ if there exists a level $p \geq 1$ such
  $\pack \circ \reduc^p(S) = \pack \circ \reduc^p(S')$ for all $S$ and $S'$ in
  $\servSet$.
\end{definition}

Table \ref{tab:reducExample} shows three proper sets, each of which projected to
a unit server. Note that the partition of a task system in proper sets depends
on the packing operation. For instance, consider $\servSet = \{ \taskdef(1, 2/3,
3\Nat^* )$, $\taskdef(2, 2/3, 3\Nat^* )$, $\taskdef(3, 1/3, 3\Nat^* )$,
$\taskdef(4, 1/3, 3\Nat^* )\}$. First, consider a packing operation $\pack$
which aggregates $(\tau_1, \tau_3)$ and $(\tau_2, \tau_4)$ into two unit servers
$S_1$ and $S_2$, then $\{\{\tau_1, \tau_3\}, \{\tau_2, \tau_4\} \}$ is the
partition of $\servSet$ into two proper sets for $\pack$. In this case, unit
servers are obtained with no reduction. Second, consider another packing
operation $\pack'$ which aggregates $(\tau_1)$, $(\tau_2)$ and $(\tau_3,
\tau_4)$ into three non-unit servers $S_1$, $S_2$ and $S_3$. Then, $\{\{\tau_1
\}, \{\tau_2\}, \{\tau_3, \tau_4\} \}$ is the partition of $\servSet$ into one
proper set for $\pack'$. In this latter case, one reduction is necessary to
obtain a unit server at level one. The correctness of the transformation,
though, does not depend on how the packing operation is implemented.

% When the specification of the packing operation is not relevant, we just write
% a proper set.  We devise the following conjecture. A proper set $\servSet$ is
% characterized by the following property: there exists a $k > 0$ such that
% $\pack \circ \reduc^k(S) = \pack \circ \reduc^k(\servSet)$, for all $S \in
% \servSet$.

%obs-old: acho que precisamos discutir este teorema.
\begin{theorem}[Reduction]\label{thm:reduction}
  Let $\reduc = \dual \,\comp\, \pack$ be a reduction and $\servSet$ be a proper
  set of EDF servers with $\util(\servSet) \in \Nat^*$ and $\pack \circ
  \reduc^p(S) = 1$ for some integer $p \geq 1$. If all servers are equipped with
  EDF, then the schedule $\sched$ of $\,\servSet$ is feasible on
  $\util(\servSet)$ processors if and only if the schedule $\,\sched'$ of $\,S'
  = \pack \comp \reduc^p(S)$ is feasible on a single virtual processor.
\end{theorem}

\begin{IEEEproof}[By transitivity between reduction levels]
  Consider the set of servers $\servSet^{(k)} = \reduc^k(\servSet)$ and its
  reduction $\reduc(\servSet^{(k)})$. By Theorem \ref{thm:reducConv},
  $\util(\servSet^{(k)}) \in \Nat$. Thus, $\pack(\servSet^{(k)})$ satisfies the
  hypothesis of Theorem \ref{thm:dualSched}. As a consequence, the schedule
  $\,\sched^{(k+1)}$ of $\,\reduc(\servSet^{(k)})$ on $|\pack(\servSet^{(k)})| -
  \util(\servSet^{(k)})$ processors is feasible if and only if the schedule
  $\bar{\sched}^{(k)}$ of $\pack(\servSet^{(k)})$ is feasible on
  $\util(\servSet^{(k)})$ processors.  As all servers in $\pack(\servSet^{(k)})$
  are EDF servers, we conclude that the schedule $\sched^{(k+1)}$ of
  $\reduc(\servSet^{(k)})$ on $|\pack(\servSet^{(k)})| - \util(\servSet^{(k)})$
  processors is feasible if and only if the schedule $\sched^{(k)}$ of
  $\servSet^{(k)}$ is feasible on $\util(\servSet^{(k)})$ processors.
\end{IEEEproof}

%obs-olhar: alterei o texto para refletir nossa última discussão
It is worth noticing that the time complexity of a reduction procedure is
polynomial. The dual operation computes for each task the utilization of its
dual, a linear time procedure. Also, since no optimality requirement is made for
implementing the packing operation, any polynomial-time heuristic applied to
pack fixed-utilization tasks/servers can be used. For example, the packing
operation can run in linear time or log-linear time, depending on the chosen
heuristic. As the following theorem shows, the time complexity of the whole
reduction procedure is dominated by the time complexity of the packing
operation.

\begin{theorem}[Reduction Complexity]\label{thm:complexity}
  The problem of scheduling $n$ fixed-utilization tasks on $m$ processors can be
  reduced to an equivalent scheduling problem on uniprocessor systems in time
  $O(f(n))$, where $f(n)$ is the time it takes to pack $n$ tasks in $m$
  processors.
\end{theorem}

\begin{IEEEproof}
  Theorem \ref{thm:reducConv} shows that a multiprocessor scheduling problem can
  be transformed into various uniprocessor scheduling problems, each of which
  formed by a proper set. Let $k$ be the largest value during a reduction
  procedure so that $\util(\reduc^k(\servSet)) = 1$, where $\servSet$ is a
  proper set. Without loss of generality, assume that $|\servSet| = n$. It must
  be shown that $k = O(f(n))$. At each step, a reduction operation is carried
  out, which costs $n$ steps for the dual operation plus $f(n)$.  Also, by Lemma
  \ref{lem:reducConv}, each time a reduction operation is applied, the number of
  tasks is divided by two. As a consequence, the time $T(n)$ to execute the
  whole reduction procedure satisfies the recurrence $T(n) = T(n/2) +
  f(n)$. Since $f(n)$ takes at least $n$ steps, the solution of this recurrence
  is $T(n) = O(f(n))$.
\end{IEEEproof}

\subsection{Illustration}\label{sec:illustration}

\begin{figure}[t]
  \centering%
  \setlength{\execWidth}{0.3\psyunit}

  \def\schedlocaxes(#1,#2){%
    \setcounter{step}{#2}%
    \multido{\nl=0+1}{\thestep}{%increments \na in step of 2 5 times
      \rput(\nl,-.25){\footnotesize \nl}%
    }%
    \multido{\i=0+1}{#1}{%
      \setcounter{proc}{\i+1}%
      \setlength{\ytmpa}{\i\psyunit+.14\psyunit}%
      \setlength{\ytmpa}{\i\psyunit}%
      \setlength{\ytmpb}{\i\psyunit-.07\psyunit}%
      \setlength{\xtmpa}{#2\psxunit-0.7\psxunit}%
      \psline[linewidth=0.5pt]{->}(0,\ytmpa)(\xtmpa,\ytmpa)%
      \multido{\nt=0+1}{#2}{
        \psline[linestyle=solid,linewidth=0.5pt](\nt,\ytmpa)(\nt,\ytmpb)%
      }%
    }%
  }%

  \setlength{\execWidth}{0.35\psxunit}%
  \def\jobleg#1#2(#3,#4,#5){%
    \setcounter{proc}{#5-1}%
    \setlength{\xtmpa}{#3\psxunit}%
    \setlength{\ytmpa}{\theproc\psyunit+0.3\psyunit}%
    \setlength{\xtmpb}{#3\psxunit+#4\psxunit}%
    \setlength{\ytmpb}{\theproc\psyunit+\execWidth+0.2\psyunit}%
    \psset{fillcolor=#1, linecolor=black,linestyle=solid,linewidth=0.5pt}%
    \psframe[linecolor=black,fillstyle=solid] (\xtmpa,\ytmpa)(\xtmpb,\ytmpb)%
    \setlength{\ytmpa}{\theproc\psyunit+0.45\psyunit}%
    \setlength{\xtmpa}{\xtmpb+0.1\psxunit}%
    \rput[l](\xtmpa,\ytmpa){#2} }%

  \def\joblocdead#1(#2,#3){%
    \setcounter{proc}{#3-1}%
    \setlength{\xtmpa}{#2\psxunit}%
    \setlength{\ytmpa}{\theproc\psyunit}%
    \setlength{\ytmpb}{\theproc\psyunit+\execWidth+0.27\psyunit}%
    \psline[linestyle=solid,linewidth=0.5pt,arrows=-*,arrowsize=4pt]
    (\xtmpa,\ytmpa)(\xtmpa,\ytmpb)%
    \ifthenelse{\equal{#1}{}}{}{%
      \setlength{\ytmpa}{\theproc\psyunit+0.74\psyunit}%
      \uput{.3em}[0](\xtmpa,\ytmpa){\vphantom{$d_j$}\footnotesize #1}}%
  }%

  \def\jobcol#1#2(#3,#4,#5){%
    \setcounter{proc}{#5-1}%
    \setlength{\xtmpa}{#3\psxunit}%
    \setlength{\ytmpa}{\theproc\psyunit}%
    \setlength{\xtmpb}{#3\psxunit+#4\psxunit}%
    \setlength{\ytmpb}{\theproc\psyunit+\execWidth}%
    \psset{fillcolor=#1, linecolor=black,linestyle=solid,linewidth=0.5pt}%
    \psframe[linecolor=black,fillstyle=solid] (\xtmpa,\ytmpa)(\xtmpb,\ytmpb)%
  }%

  \newcommand{\colA}{blue!50!white} \newcommand{\colB}{blue!90!white}
  \newcommand{\colAB}{blue!94!black}

  \newcommand{\colC}{red!50!white} \newcommand{\colD}{red!90!white}
  \newcommand{\colCD}{red!94!black}

  \newcommand{\colE}{yellow!97!black!80!white}
  \newcommand{\colEE}{yellow!97!black}

  % \setbeamertemplate{background canvas}[vertical
  % shading][bottom=white!10,top=structure.fg!20]%
  \psset{xunit=1.2cm, yunit=0.8cm}
  \begin{pspicture*}(-0.5,0)(8.5,2.1)%
    \jobleg{\colA}{\footnotesize $S_1, S_1^*$}(0, 0.2, 2)%
    \jobleg{\colB}{\footnotesize $S_2, S_2^*$}(1.2, 0.2, 2)%
    \jobleg{\colC}{\footnotesize $S_3, S_3^*$}(2.4, 0.2, 2)%
    \jobleg{\colD}{\footnotesize $S_4, S_4^*$}(3.6, 0.2, 2)%
    \jobleg{\colE}{\footnotesize $S_5, S_5^*$}(4.8, 0.2, 2)%

    \jobleg{\colAB}{\footnotesize $\dualMap \circ \serv \{ S_1^*, S_2^* \}$}(0,
    0.2, 1)%
    \jobleg{\colCD}{\footnotesize $\dualMap \circ \serv \{ S_3^*, S_4^*
      \}$}(2.4, 0.2, 1)%
    \jobleg{\colEE}{\footnotesize $\dualMap \circ \serv \{ S_5^* \}$}(4.8, 0.2,
    1)%
  \end{pspicture*}
  \psset{xunit=1.2cm, yunit=1cm}
  \begin{pspicture*}(-0.8,-0.4)(8.5,6.)%
    \schedlocaxes(6,7)
    % Task 1
    \rput[l](-0.74, 4.15){$\Sigma_0$}%
    \psbezier[linewidth=0.5pt]{-}(-0.3,4.2)(-0.1,4.4)(-0.3,5.3)(-0.1,5.5)
    \psbezier[linewidth=0.5pt]{-}(-0.3,4.2)(-0.1,4.)(-0.3,3.1)(-0.1,2.9)
    \psbezier[linewidth=0.5pt]{-}(-0.3,1.7)(-0.1,1.9)(-0.3,2.3)(-0.1,2.5)
    \psbezier[linewidth=0.5pt]{-}(-0.3,1.7)(-0.1,1.5)(-0.3,1.1)(-0.1,0.9)
    \psbezier[linewidth=0.5pt]{-}(-0.3,0.2)(-0.1,0.3)(-0.3,0.45)(-0.1,0.5)
    \psbezier[linewidth=0.5pt]{-}(-0.3,0.2)(-0.1,0.1)(-0.3,-0.05)(-0.1,-0.1)
    % Proc 1
    \jobcol{\colE}(0, 1.2, 6)%
    \jobcol{\colD}(1.2, 1.8, 6)%
    \jobcol{\colC}(3, 0.2, 6)%
    \jobcol{\colE}(3.2, 0.2, 6)%
    \jobcol{\colA}(3.4, 1, 6)%
    \jobcol{\colD}(4.4, 0.4, 6)%
    \jobcol{\colE}(4.8, 1.2, 6)%
    % Proc 2
    \jobcol{\colC}(0, 1.8, 5)%
    \jobcol{\colB}(1.8, 0.2, 5)%
    \jobcol{\colE}(2, 1, 5)%
    \jobcol{\colD}(3, 0.6, 5)%
    \jobcol{\colC}(3.6, 0.4, 5)%
    \jobcol{\colC}(4, 1.2, 5)%
    \jobcol{\colD}(5.2, 0.8, 5)%
    % Proc 3
    \jobcol{\colA}(0, 2.6, 4)%
    \jobcol{\colB}(2.6, 3.4, 4)%
    \joblocdead{1}(2,6)%
    \joblocdead{2}(3,6)%
    \joblocdead{1}(4,6)%
    \joblocdead{2}(6,5)%
    \joblocdead{3}(4,5)%
    \joblocdead{4}(6,4)%
    % \schedlocaxes(2,7)%
    \rput[l](-0.74, 1.65){$\Sigma_1$}
    % Proc 2
    \jobcol{\colB}(0, 1.8, 3)%
    \jobcol{\colC}(1.8, 1.2, 3)%
    \jobcol{\colE}(3, 0.2, 3)%
    \jobcol{\colC}(3.2, 0.4, 3)%
    \jobcol{\colD}(3.6, 0.8, 3)%
    \jobcol{\colA}(4.4, 1.6, 3)%
    % Proc 1
    \jobcol{\colD}(0, 1.2, 2)%
    \jobcol{\colE}(1.2, 0.8, 2)%
    \jobcol{\colB}(2, 0.6, 2)%
    \jobcol{\colA}(2.6, 0.8, 2)%
    \jobcol{\colE}(3.4, 0.6, 2)%
    \jobcol{\colE}(4, 0.8, 2)%
    \jobcol{\colD}(4.8, 0.4, 2)%
    \jobcol{\colC}(5.2, 0.8, 2)%
    % \schedlocaxes(1,7)%
    \rput[l](-0.74, 0.15){$\Sigma_2$}
    % Task 1
    \jobcol{\colEE}(0, 1.2, 1)%
    \jobcol{\colCD}(1.2, 0.6, 1)%
    \jobcol{\colAB}(1.8, 0.2, 1)%
    \jobcol{\colEE}(2, 1, 1)%
    \jobcol{\colCD}(3, 0.2, 1)%
    \jobcol{\colEE}(3.2, 0.2, 1)%
    \jobcol{\colAB}(3.4, 1, 1)%
    \jobcol{\colCD}(4.4, 0.4, 1)%
    \jobcol{\colEE}(4.8, 1.2, 1)%

    \joblocdead{}(2,1)%
    \joblocdead{}(3,1)%
    \joblocdead{}(4,1)%
    \joblocdead{}(6,1)%
    \psset{linewidth=0.5pt,linestyle=dashed,dash=2pt 3pt}
    \psline(1.2,-0.1)(1.2,5.6) \psline(1.8,-0.1)(1.8,5.6) \psline(2,-0.1)(2,5.6)
    \psline(2.6,-0.1)(2.6,5.6) \psline(3,-0.1)(3,5.6) \psline(3.2,-0.1)(3.2,5.6)
    \psline(3.4,-0.1)(3.4,5.6) \psline(3.6,-0.1)(3.6,5.6)
    \psline(4.,-0.1)(4.,5.6) \psline(4.4,-0.1)(4.4,5.6)
    \psline(4.8,-0.1)(4.8,5.6) \psline(5.2,-0.1)(5.2,5.6)
  \end{pspicture*}
  \caption{$\servSet = \{ S_1, S_2, S_3, S_4, S_5\}$ with $S_1 =
    \serv(\taskdef(1, 3/5, 2\Nat^* ))$, $S_2 = \serv(\taskdef(2, 3/5, 3\Nat^*
    ))$, $S_3 = \serv(\taskdef(3, 3/5, 4\Nat^* ))$, $S_4 = \serv(\taskdef(4,
    3/5, 6\Nat^* ))$ and $S_5 = \serv(\taskdef(5, 3/5, 12\Nat^* ))$. $\sched_0$,
    $\sched_1$ and $\sched_2$ are the schedule on three processors, two virtual
    processors and one virtual processor of $\servSet$, $\dual(\servSet)$ and
    $\reduc \circ \dual(\servSet)$, respectively.}
  \label{fig:dualSchedComplex}
\end{figure}
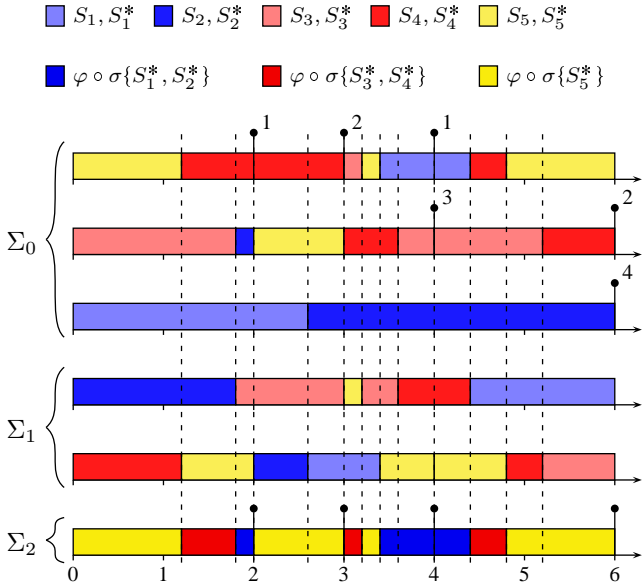

Figure \ref{fig:dualSchedComplex} shows an illustrative example produced by
simulation with a task set which requires two reduction levels to be
scheduled. Observe, for instance, that when $\dual(\serv\{ S_3^*, S_4^* \})$ is
executing in $\sched_2$, then both $S_3^*$ and $S_4^*$ do not execute in
$\sched_1$, and both $S_3$ and $S_4$ execute in real schedule $\sched_0$. On the
other hand, when $\dual(\serv\{ S_3^*, S_4^* \})$ does not execute in
$\sched_2$, then either -- $S_3^*$ and $S_4$ -- or exclusive -- $S_4^*$ and
$S_3$ -- executes in $\sched_1$ and $\sched_0$, respectively.

% The following heuristic was used for the packing operation $\pack$ of a task
% set $\servSet$. Picks-up the highest utilization task in $\servSet$ and
% allocates it to the least utilization server already created, if
% possible. Otherwise, creates a new zero-utilization server and fill it with
% the current task.

%\cleardoublepage
\section{Assessment}\label{sec:evaluation}

We have carried out intensive simulation to evaluate the proposed approach. We
generated one thousand random task sets with $n$ tasks each, $n = 17, 18, 20,
22, \ldots, 64$. Hence a total of $24$ thousands task sets were generated. Each
task set fully utilizes a system with $16$ processors.  Although other
utilization values were considered, they are not shown here since they presented
similar result patterns. The utilization of each task was generated following
the procedure described in \cite{Emberson10a}, using the aleatory task generator
by \cite{Emberson10b}. Task periods were generated according to a uniform
distribution in the interval $[5,100]$.

Two parameters were observed during the simulation, the number of reduction
levels and the number of preemption points occurring on the real multiprocessor
system. Job completion is not considered as a preemption point. The results were
obtained implementing the packing operation using the decreasing worst-fit
packing heuristic.

\begin{figure}[t]
  %\centering%
  \includegraphics{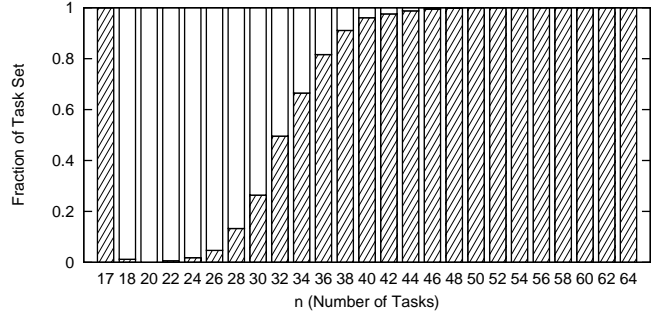}
  \caption{Fraction of task sets which requires 1 (crosshatch box) and 2 (empty
    box) reduction levels. 1000 task sets were generated for each point.}
  \label{fig:level}
\end{figure}

Figure \ref{fig:level} shows the number of reduction levels. It is interesting
to note that none of the task sets generated required more than two reduction
levels. For $17$ tasks, only one level was necessary. This situation,
illustrated in Figure \ref{fig:dualSchedEx}, is a special case of Theorem
\ref{thm:dualSched}. One or two levels were used for $n$ in $[18,48]$. For
systems with more than $48$ tasks, the average task utilization is low. This
means that the utilization of each server after performing the first packing
operation is probably close to one, decreasing the number of necessary
reductions.

\begin{figure}[t]
  %\centering%
  \includegraphics{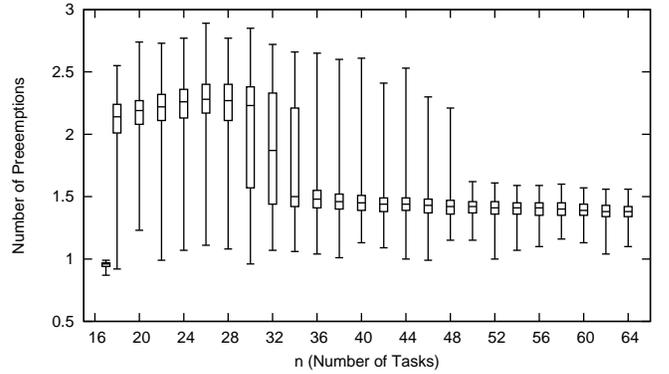}
  \caption{Distributions of the average number of preemptions per job, their
    quartiles, and their minimum and maximum values.}
  \label{fig:ratio}
\end{figure}

The box-plot shown in Figure \ref{fig:ratio} depicts the distribution of
preemption points as a function of the number of tasks. The number of
preemptions is expected to increase with the number of levels and with the
number of tasks packed into each server. This behavior is observed in the
figure, which shows that the number of levels has a greater impact. Indeed, the
median regarding scenarios for $n$ in $[52,64]$ is below $1.5$ and for those
scenarios each server is likely to contain a higher number of tasks. Further,
observe that the maximum value observed was $2.8$ preemption points per job on
average, which illustrates a good characteristic of the proposed approach.

\section{Related Work}\label{sec:relatedWork}

Solutions to the real-time multiprocessor scheduling problem can be
characterized according to the way task migration is controlled. Approaches
which do not impose any restriction on task migration are usually called global
scheduling. Those that do not allow task migration are known as partition
scheduling. Although partition-based approaches make it possible using the
results for uniprocessor scheduling straightforwardly, they are not applicable
for task sets which cannot be correctly partitioned. On the other hand, global
scheduling can provide effective use of a multiprocessor architecture although
with possibly higher implementation overhead.

There exist a few optimal global scheduling approaches for the PPID model. If
all tasks share the same deadline, it has been shown that the system can be
optimally scheduled with a very low implementation cost
\cite{McNaughton59}. Removing this restriction on task deadlines, optimality can
be achieved by approaches that approximate the theoretical fluid model,
according to which all tasks execute at the steady rate proportional to their
utilization \cite{Baruah96}. However, this fluid approach has the main drawback
that it potentially generates an arbitrary large number of preemptions.

Executing all tasks at a steady rate is also the goal of other approaches
\cite{Cho06,Funaoka08}. Instead of breaking all task in fixed-size quantum
subtasks, such approaches define scheduling windows, called T-L planes, which
are intervals between consecutive task deadlines.
% During a node $N$ of length $\delta$, each active task of the periodic taskset
% executes for $\,\delta C / T$. Whenever the laxity of a task reaches zero, a
% secondary scheduling instant is created and the zero laxity task is scheduled
% to execute immediately, until its deadline.
The T-L plane approach has been extended recently to accommodate more general
task models \cite{Funk10}. Although the number of generated preemptions has
shown to be bounded within each T-L plane, the number of T-L planes can be
arbitrarily high for some task sets.

%With respect to the constrained sporadic job model, in which $T_i$ is the
%minimal inter-arrival time between two jobs of the same task (sporadic) and the
%relative deadline of a task can be smaller than $T_i$ (constrained), it is known from \cite{Sahni79, Hong88, Dertouzos89, Fischer09} that there exists no on-line
%optimal algorithm to schedule jobs on a multiprocessor system.  Hence, Cho and
%al. \cite{Cho02} introduce the weaker notion of suboptimality. A preemptive
%algorithm is suboptimal if it can successfully schedule any feasible set of
%ready jobs, where a ready job at time $t$ is a job that has been released at or
%before $t$. It is shown in \cite{Cho02} that neither the EDF nor Earliest
%Deadline Zero Laxity (EDZL) algorithms are suboptimal when used to schedule jobs
%on more than one processor. On the other hand, the Least Laxity First algorithm
%is suboptimal \cite{Dertouzos89}. Since this latter algorithm generates a high
%number of preemptions for some task sets, enhanced schemes are needed to turn it
%suitable \cite{Hildebrandt99}. Also, such schemes do not lead to known optimal
%algorithm at the current time.

Other approaches which control task migration have been proposed
\cite{Andersson08, Easwaran09, Kato09, Massa10}. They have been called
semi-partition approaches. The basic idea is to partition some tasks into
disjunct subsets. Each subset is allocated to processors off-line, similar to
the partition-based approaches. Some tasks are allowed to be allocated to more
than one processor and their migration is controlled at run-time. Usually, these
approaches present a trade-off between implementation overhead and achievable
utilization, and optimality can be obtained if preemption overhead is not
bounded.

The approach presented in this paper lie in between partition and global
approaches. It does not assign tasks to processors but to servers and optimality
is achieved with low preemption cost. Task migration is allowed but is
controlled by the rules of both the servers and the virtual schedule. Also, as
the scheduling problem is reduced from multiprocessor to uniprocessor, well
known results for uniprocessor systems can be used. Indeed, optimality for
fixed-utilization task set on multiprocessor is obtained by using an optimal
uniprocessor scheduler, maintaining a low preemption cost per task.

% The approach described in this paper achieves optimality without assuming that
% tasks have the same deadline. However, only a few preemption points per job are
% generated.

It has recently been noted that if a set with $m+1$ tasks have their total
utilization exactly equal to $m$, then a feasible schedule of these tasks on $m$
identical processors can be produced \cite{Levin09}. The approach described here
generalizes this result. The use of servers was a key tool to achieve this
generalization. The concept of task servers has been extensively used to provide
a mechanism to schedule soft tasks \cite{liu00}, for which timing attributes
like period or execution time are not known a priori. There are some server
mechanisms for uniprocessor systems which share some similarities with one
presented here \cite{DLS97,SB96}. To the best of our knowledge the server
mechanism presented here is the first one designed with the purposes of solving
the real-time multiprocessor scheduling problem.

\section{Conclusion}\label{sec:conclusion}

An approach to scheduling a set of tasks on a set of identical multiprocessors
has been described. The novelty of the approach lies in transforming the
multiprocessor scheduling problem into an equivalent uniprocessor
one. Simulation results have shown that only a few preemption points per job on
average are generated.

The results presented here have both practical and theoretical
implications. Implementing the described approach on actual multiprocessor
architectures is among the practical issues to be explored. Theoretical aspects
are related to relaxing the assumed task model, \textit{e.g.} sporadic tasks
with constrained deadlines. Further, interesting questions about introducing new
aspects in the multiprocessor schedule via the virtual uniprocessor schedule can
be raised. For example, one may be interested in considering aspects such as
fault tolerance, energy consumption or adaptability. These issues are certainly
a fertile research field to be explored.

%To do so,
%a server mechanism and two basic operations have been defined. Servers provide %a
%powerful scheduling abstraction and are responsible for aggregating tasks (or
%servers) and generating necessary preemption points. The dual and packing
%operations are used to reduce, respectively, the utilization of the system and
%the number of tasks to be scheduled. 

%\newpage
\bibliographystyle{IEEEtran}
\bibliography{bib}

\begin{thebibliography}{10}
\providecommand{\url}[1]{#1}
\csname url@samestyle\endcsname
\providecommand{\newblock}{\relax}
\providecommand{\bibinfo}[2]{#2}
\providecommand{\BIBentrySTDinterwordspacing}{\spaceskip=0pt\relax}
\providecommand{\BIBentryALTinterwordstretchfactor}{4}
\providecommand{\BIBentryALTinterwordspacing}{\spaceskip=\fontdimen2\font plus
\BIBentryALTinterwordstretchfactor\fontdimen3\font minus
  \fontdimen4\font\relax}
\providecommand{\BIBforeignlanguage}[2]{{%
\expandafter\ifx\csname l@#1\endcsname\relax
\typeout{** WARNING: IEEEtran.bst: No hyphenation pattern has been}%
\typeout{** loaded for the language `#1'. Using the pattern for}%
\typeout{** the default language instead.}%
\else
\language=\csname l@#1\endcsname
\fi
#2}}
\providecommand{\BIBdecl}{\relax}
\BIBdecl

\bibitem{McNaughton59}
R.~McNaughton, ``Scheduling with deadlines and loss functions,''
  \emph{Management Science}, vol.~6, no.~1, pp. 1--12, 1959.

\bibitem{Baruah96}
S.~Baruah, N.~K. Cohen, C.~G. Plaxton, and D.~A. Varvel, ``Proportionate
  progress: A notion of fairness in resource allocation,'' \emph{Algorithmica},
  vol.~15, no.~6, pp. 600--625, 1996.

\bibitem{Cho06}
H.~Cho, B.~Ravindran, and E.~D. Jensen, ``An optimal real-time scheduling
  algorithm for multiprocessors,'' in \emph{27th IEEE Real-Time Systems Symp.},
  2006, pp. 101--110.

\bibitem{Levin2010}
G.~Levin, S.~Funk, C.~Sadowski, I.~Pye, and S.~Brandt, ``{DP-FAIR}: A simple
  model for understanding optimal multiprocessor scheduling,'' in
  \emph{Euromicro Conf. on Real-Time Systems}, 2010, pp. 3--13.

\bibitem{Andersson08}
B.~Andersson, K.~Bletsas, and S.~Baruah, ``Scheduling arbitrary-deadline
  sporadic task systems on multiprocessors,'' in \emph{29th IEEE Real-Time
  Systems Symp.}, 2008, pp. 385--394.

\bibitem{Massa10}
E.~Massa and G.~Lima, ``A bandwidth reservation strategy for multiprocessor
  real-time scheduling,'' in \emph{16th IEEE Real-Time and Embedded Technology
  and Applications Symp.}, april 2010, pp. 175 --183.

\bibitem{Andersson08b}
B.~Andersson and K.~Bletsas, ``Sporadic multiprocessor scheduling with few
  preemptions,'' in \emph{20th Euromicro Conf. on Real-Time Systems}, July
  2008, pp. 243--252.

\bibitem{Bletsas09}
K.~Bletsas and B.~Andersson, ``Notional processors: An approach for
  multiprocessor scheduling,'' in \emph{15th IEEE Real-Time and Embedded
  Technology and Applications Symp.}, April 2009, pp. 3--12.

\bibitem{Liu73}
C.~L. Liu and J.~W. Layland, ``Scheduling algorithms for multiprogram in a hard
  real-time environment,'' \emph{Journal of ACM}, vol.~20, no.~1, pp. 40--61,
  1973.

\bibitem{Emberson10a}
P.~Emberson, R.~Stafford, and R.~I. Davis, ``Techniques for the synthesis of
  multiprocessor tasksets,'' in \emph{Proc. of 1st Int. Workshop on Analysis
  Tools and Methodologies for Embedded and Real-time Systems (WATERS 2010)},
  2010, pp. 6--11.

\bibitem{Emberson10b}
------, ``A taskset generator for experiments with real-time task sets,''
  {http://retis.sssup.it/waters2010/data/taskgen-0.1.tar.gz}, Jan. 2011.

\bibitem{Funaoka08}
K.~Funaoka, S.~Kato, and N.~Yamasaki, ``Work-conserving optimal real-time
  scheduling on multiprocessors,'' in \emph{20th Euromicro Conf. on Real-Time
  Systems}, 2008, pp. 13--22.

\bibitem{Funk10}
S.~Funk, ``An optimal multiprocessor algorithm for sporadic task sets with
  unconstrained deadlines,'' \emph{Real-Time Systems}, vol.~46, pp. 332--359,
  2010.

\bibitem{Easwaran09}
A.~Easwaran, I.~Shin, and I.~Lee, ``Optimal virtual cluster-based
  multiprocessor scheduling,'' \emph{Real-Time Syst.}, vol.~43, no.~1, pp.
  25--59, 2009.

\bibitem{Kato09}
S.~Kato, N.~Yamasaki, and Y.~Ishikawa, ``Semi-partitioned scheduling of
  sporadic task systems on multiprocessors,'' in \emph{21st Euromicro Conf. on
  Real-Time Systems}, 2009, pp. 249--258.

\bibitem{Levin09}
G.~Levin, C.~Sadowski, I.~Pye, and S.~Brandt, ``{S{\small N}S}: A simple model
  for understanding optimal hard real-time multi-processor scheduling,'' Univ.
  of California, Tech. Rep., 2009.

\bibitem{liu00}
J.~W.~S. Liu, \emph{Real-Time Systems}.\hskip 1em plus 0.5em minus 0.4em\relax
  Prentice-Hall, 2000.

\bibitem{DLS97}
Z.~Deng, J.~W.-S. Liu, and J.~Sun, ``A scheme for scheduling hard real-time
  applications in open system environment,'' in \emph{9th Euromicro Workshop on
  Real-Time Systems}, 1997, pp. 191--199.

\bibitem{SB96}
M.~Spuri and G.~Buttazzo, ``Scheduling aperiodic tasks in dynamic priority
  systems,'' \emph{Real Time Systems}, vol.~10, no.~2, pp. 179--210, 1996.

\end{thebibliography}

\end{document}